\documentclass[sigconf, screen]{acmart}
\usepackage{amsmath,amssymb,amsfonts}
\usepackage{algorithmic}
\usepackage{graphicx}
\usepackage{textcomp}
\usepackage{xspace}
\usepackage{algorithm}
\usepackage{algorithmic}
\usepackage{amssymb}
\usepackage{amsmath}
\usepackage{amsthm}
\usepackage{subfigure}
\usepackage{placeins}
\usepackage{url}
\usepackage{hyperref}
\usepackage{multirow}

\newcommand{\name}{LogPlayer\xspace}
\newcommand\x{0.945}

\newcommand\diagramScale{0.88}
\newcommand\archScale{0.65}

\begin{document}
 \title{\name: Fault-tolerant Exactly-once Delivery using gRPC Asynchronous Streaming}


\author{Mohammad Roohitavaf}
\affiliation{%
  \institution{eBay Inc.}
  \city{San Jose}
  \state{CA}
   \country{USA}
}
\email{mroohitavaf@ebay.com}

\author{Kun Ren}
\affiliation{%
  \institution{eBay Inc.}
  \city{San Jose}
  \state{CA}
   \country{USA}
}
\email{kuren@ebay.com}

\author{Guogen Zhang}
\affiliation{%
  \institution{Netskope Inc.}
  \city{Santa Clara}
  \state{CA}
   \country{USA}
}
\authornote{Work done while the author was at eBay Inc.}
\email{gzhang@netskope.com}

\author{Sami Ben-Romdhane}
\affiliation{%
  \institution{eBay Inc.}
  \city{San Jose}
  \state{CA}
   \country{USA}
}
\email{sbenromdhane@ebay.com}

\begin{abstract}
In this paper, we present the design of our \name that is a component responsible for fault-tolerant delivery of transactional mutations recorded on a WAL to the backend storage shards.  \name relies on gRPC for asynchronous streaming. However, the design provided in this paper can be used with other asynchronous streaming platforms. We model check the correctness of \name by TLA+. In particular, our TLA+ specification shows that \name guarantees in-order exactly-once delivery of WAL entries to the storage shards, even in the presence of shards or \name failures. Our experiments show \name is capable of efficient delivery with sub-millisecond latency, and it is significantly more efficient than Apache Kafka for designing a WAL system with exactly-once guarantee.
\end{abstract}


\maketitle
\section{Introduction}
\label{sec:intro}

Write-ahead logging \cite{gray} is one of the most popular approaches for providing atomicity and durability for transactions. A database management system usually appends the mutations of a transaction to a Write-Ahead Log (WAL), before applying the mutations to the actual data. 
For a single-node database, we can maintain the WAL as an append-only file on the disk. For a distributed database, however, this approach makes the machine maintaining the WAL a single-point-of-failure. To increase the availability, we can maintain our WAL as a \textit{replicated} log hosted by a set of machines. Using distributed consensus algorithms such as raft \cite{raft}, we can guarantee that all machines maintain the same WAL entries with the same order, and writes to the WAL are durable even in case of failure of some nodes. 
  Once we have our highly available WAL, we need a mechanism to apply WAL entries to the backend storage shards. It is crucial to make sure that each WAL entry is applied to the data exactly one time-- a semantics that is known as \textit{exactly-once}. If some of the mutations are missed or applied more than one time, the correctness of the applications may be violated. For example, consider a transaction that adds an item to a customer's shopping cart. For this transaction, either missing the item or adding it to the shopping cart more than one time is unacceptable.
 
 In this paper, we present the design of \name, a component responsible for pushing mutations of transactions written to a WAL to backend storage shards with the exactly-once delivery guarantee. \name uses asynchronous streaming to efficiently send updates to storage shards without being affected by failed or degraded shards.  \name relies on gRPC \cite{grpc} for asynchronous streaming. However, the design provided in this paper can be used with any other asynchronous streaming platform that utilizes the notion of the completion queue. 
 We provide a TLA+ \cite{tla} specification, in PlusCal language, for model checking the correctness of our \name. This model proves the algorithms provided in this paper guarantee exactly-once delivery even in the presence of shard or \name failures. 


 Deterministic databases or variants\cite{thomson2012calvin, ren2019slog, faunadb} have shown promise to remove expensive commit protocols in scalable distributed deployments, and enable higher amounts of transactional throughput and concurrency. \name is among critical components for a deterministic database system, as it facilitates pushing mutations of the committed transactions to the database shards. Currently, \name is used in an architecture for providing strictly serializable transactions across microservices called GRIT developed for eBay's data platform \cite{grit}. This architecture has been also used for developing a transactional backend storage for JanusGraph~\cite{janusgraph}. 
  
Our experiments with the C++ implementation of \name shows it can efficiently deliver log entries to the storage shards. In all of our experiments, the median and average \name delay remain less than 1 millisecond. We show our system is significantly faster than a WAL system created using Apache Kafka \cite{kafka} with the same quality of service. For instance, for transactions with 10~KB payload spanning 10 shards, our system,  on average, is 4 times faster than a similar system with Apache Kafka. The advantage of our system is greater for the tail latency, as we observed our system is 6 times faster than the alternative design with Apache Kafka regarding the 99th percentile of the delay of delivering WAL entries to the storage shards.  


The rest of paper is organized as follows: In Section \ref{sec:design}, we provide the architecture and algorithms of \name. Section \ref{sec:correctness} provides the TLA+ specification. In Section \ref{sec:wal}, we discuss how \name can be configured with LogStore to form a highly available WAL system. Section \ref{sec:related} reviews the related work and provides an alternative design with Apache Kafka. We provide experimental results in Section \ref{sec:results}. Finally, Section \ref{sec:con} concludes the paper.

\section{\name Design}
\label{sec:design}
In this section, we focus on the design of \name. We first provide the overall architecture and then, explain the components in more detail. 

\subsection{Architecture}
\label{sec:arch}
\begin{figure} [h]
\begin{center}
\includegraphics[width=0.8\columnwidth]{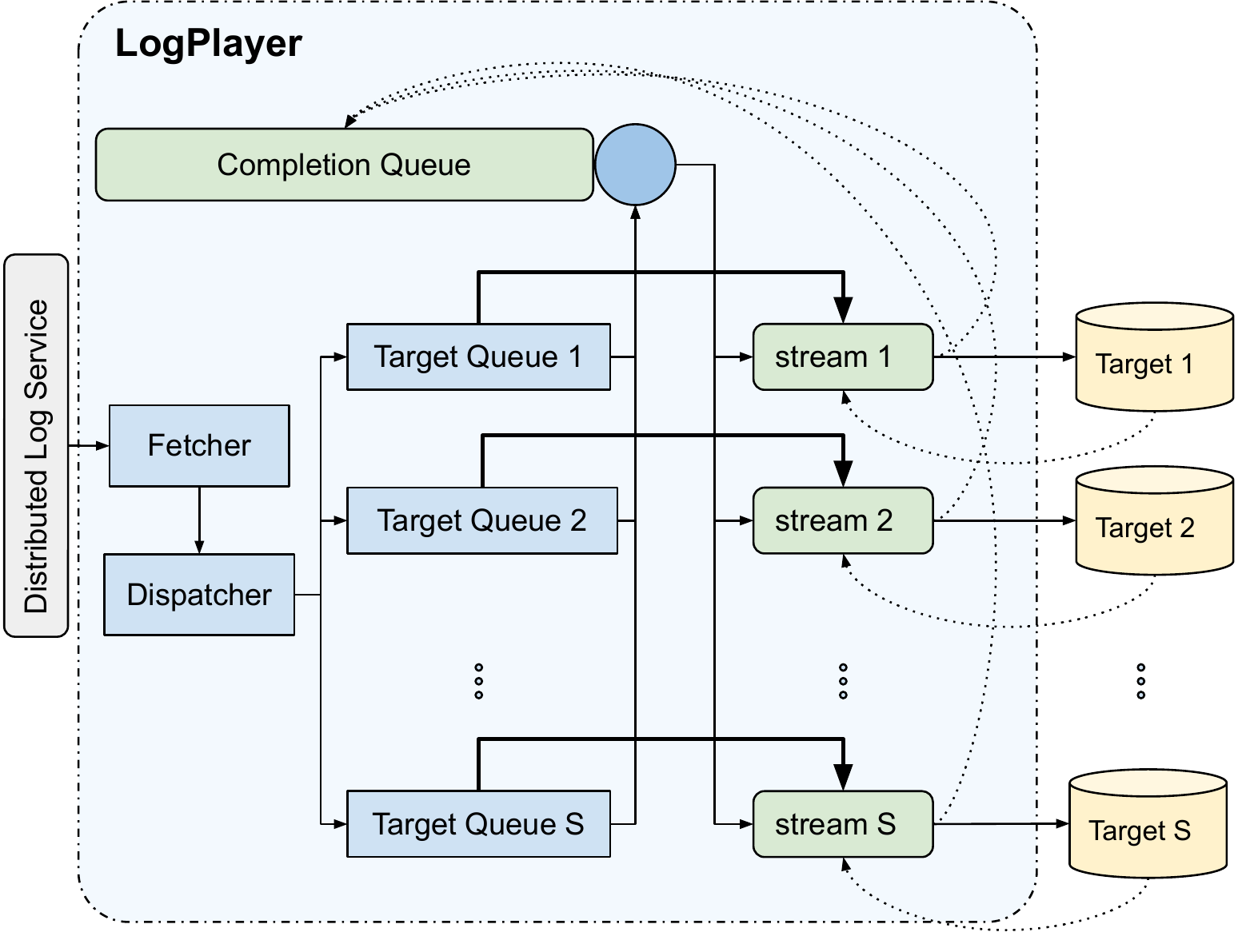}
\caption{Architecture of \name}
\label{fig:arch}
\end{center}
\end{figure}

Figure \ref{fig:arch} shows the architecture of \name. The job of \name is reading the entries of a log service, processing, and sending them to the right destinations. We call each destination a \textit{target}. The communication with the log service is done by the fetcher component. This component is responsible for feeding the \name with a stream of log entries. Using a fetcher, \name creates a stream, called \textit{main stream}, for reading log entries and delivering them to the current healthy targets. This stream never finishes and is supposed to run for the entire lifetime of a \name execution. In addition to this ongoing stream, ad hoc \textit{recovery streams} may also be created for reading only limited ranges of entries missed by failed targets after they come back. Once the fetcher receives a new entry from the log service, it calls the dispatcher. The dispatcher processes the log entry and adds proper messages to the target queues. 
Keeping a separate queue for each target prevents a slow or failed target from slowing the delivery to the other targets. 

\name constantly checks the health of the targets and stops adding messages to a target queue, once it finds out the target has crashed/disconnected. Thus, a target misses the entries that are dispatched while the target is down/disconnected. When the target comes back, \name creates a recovery stream that is responsible for reading entries missed by the target. The target joins the main stream again when the catchup process is completed. We explain the details of this process in Section \ref{sec:targetQueue}. 
To communicate with the targets, \name relies on gRPC~\cite{grpc} for streaming with FIFO guarantee. To each target, \name maintains a bidirectional asynchronous streaming channel. An asynchronous streaming channel in gRPC works with a completion queue. Specifically, we assign a tag to each read/write operation. The operation does not block. Instead, when gRPC is done with the operation, the tag we assigned to it shows up on the completion queue. We can have multiple completion queues each assigned to a set of streams. However, since processing the completion queue is very lightweight (as we will see Section \ref{sec:cq}), we believe having only one completion queue with one processing thread for all streams is enough. Alternatively, one could use \textit{synchronous} streaming. Using a bidirectional synchronous streaming channel, the sender can keep writing to the stream in one thread, and read the responses concurrently in another thread. Since in the synchronous streaming, read operations are blocking, this approach needs $n$ threads to read the responses from $n$ targets which is not desired. That is why we preferred asynchronous streaming over synchronous streaming. 

\name guarantees in-order exactly-once delivery of log entries to the targets, i.e., targets do not need to check the entries that they receive from \name to make sure that they are not out-of-order or duplicate. The targets, however, must make sure consuming an entry happens atomically together with durably storing the index of the entry. In particular, a target must be able to retrieve the index of the last entry that it has consumed, even after recovering from a crash. In Section~\ref{sec:recovery}, we explain how this index is used to resume delivering entries to a failed target after it comes back.
Next, we focus on the details of the components. 

\subsection{Target Queue}
\label{sec:targetQueue}
\begin{figure} [h]
\begin{center}
\includegraphics[width=\archScale\columnwidth]{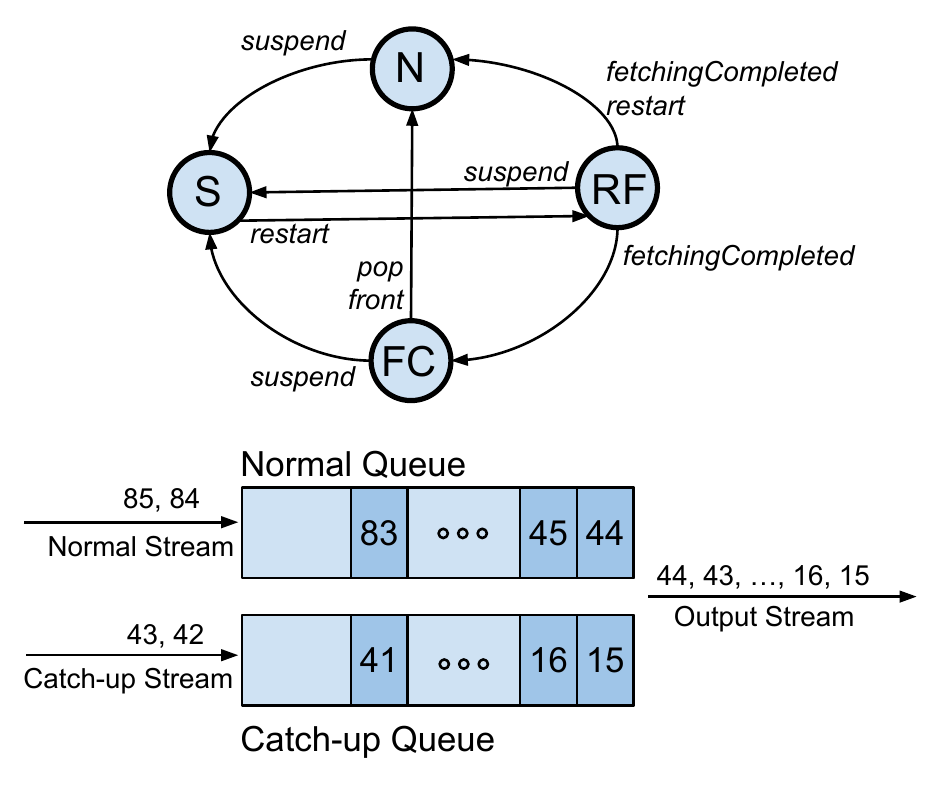}
\caption{Internal queues and states of a target queue}
\label{fig:target_queue}
\end{center}
\end{figure}

A target queue is responsible for maintaining the entries destined at a target. In addition to the typical queue interface, the target queue provides necessary functionalities for batching and recovery. Each target queue has two internal typical queues. One queue is called \textit{normal queue} that maintains the entries of the main stream, and the other is called \textit{catchup queue} that maintains the entries of the recovery stream. 
A target queue has four states shown in Figure~\ref{fig:target_queue}. The Normal ($\texttt{N}$) state means that the target which this queue belongs to is up and no recovery is in process for that target. The Recovery Fetching ($\texttt{RF}$) state means that the target is up, but a recovery is in process for it and fetching from log service is not completed (i.e., there exists an entry of the main stream missed by the target that has not been yet pushed to the queue). The Fetching Completed ($\texttt{FC}$)  means that the target is up, recovery is in process, and fetching from log service is completed. Finally, the Suspended ($\texttt{S}$) state means the target is down/disconnected. 

The basic behavior of the target queue is shown in Algorithm \ref{alg:targetQueue1}. To push a new entry, we have to specify whether the entry belongs to the main stream or a  recovery stream. If the entry belongs to the main stream, the target queue adds it to its normal queue. Otherwise, it adds the entry to its catch-up queue. We assign a \textit{term} to each target. The term of a target is initialized to 1, and it increments by one every time the target crashes/disconnects and comes back. When we are pushing an entry to a target queue, we have to specify that the entry is intended to be sent in which term. If the term is expired, the target queue drops the entry. This mechanism is designed to prevent delivering duplicate entries to a target due to failures. The target queue also does not push any entry, when it is in state $\texttt{S}$. 
To access the front entry of the queue, the target queue checks its state. Being in $\texttt{RF}$ or $\texttt{FC}$ means that there are still entries of the main stream missed by the target that have to be sent to the target. Thus, if the queue is in $\texttt{RF}$ or $\texttt{FC}$, and catchup queue is not empty, the target queue returns the front of its catchup queue. If the queue is in the normal state and the normal queue is not empty, it returns the front of its normal queue. In any other situation, it returns $\perp$ indicating that the queue is not ready to provide the next entry. 
Similarly, the popping behavior depends on the state of the queue. If queue is in $\texttt{RF}$ or $\texttt{FC}$, it pops the catchup queue. If the queue is in $\texttt{FC}$ and the catchup queue is empty after this pop, the queue transitions to the normal state, as all missed entries are popped from the queue and target is ready to receive entries pushed by the main stream. If the queue is in the normal state, it simply pops its normal queue. 

Once a target is down/disconnected, the health checker calls the $suspend$ function of the corresponding target queue. This routine clears the normal and catchup queues and sets the state of the queue to $\texttt{S}$. As mentioned above and we will explain in more detail in Section \ref{sec:recovery}, once a failed target comes back, a recovery stream may be created to provide it with entries that it has missed. When the recovery fetcher is done with pushing entries missed by the target, it calls the $fetchingCompleted$ routine. If the queue is in $\texttt{RF}$ and term has not expired, $fetchingCompleted$ changes the state of the queue depending on the catchup queue. If the catchup queue is empty, it means all missed entries have been already sent to the target. Thus, it changes the state to $\texttt{N}$ and calls $sendNext$ to send the next entry to the target. Otherwise, it changes the state to $\texttt{FC}$, which means fetching the missed entries is done, but sending them to the target is not finished yet. 

To reduce the number of gRPC calls and consequently the number of entries in the completion queue, we send the entries in batches. Algorithm \ref{alg:targetQueue2} shows the batching behavior of the target queue. To create a new batch, we keep accessing the front entry of the queue as explained above, until the front entry is not available or batch size reaches the given maximum value. After sending a batch, we add it to a map called $popped$. The target queue does not delete the memory allocated to the popped batches unless being explicitly called to do that. In Section \ref{sec:cq}, we explain when popping and deleting batches are called for target queues. 

\begin{algorithm} [h]
{
\small
\caption{Target Queue - part 1}
\label{alg:targetQueue1}
\begin{algorithmic} [1]

\STATE \textbf{push} ($e$, $isNormal$, $term$)
\STATE \hspace{3mm} lock ($L_q$)
\STATE \hspace{3mm} \textbf{if} ($state \neq$  $\texttt{S}$ $\wedge   (isNormal \vee term = current\_term)$)
\STATE \hspace{6mm} \label{line:pushToNormal} \textbf{if} ($isNormal$) $normal$.push($e$)
\STATE \hspace{6mm} \textbf{else} $catchup$.push($e$)
\STATE \hspace{3mm} unlock ($L_q$)

\vspace{2mm}

\STATE \textbf{front}
\STATE \hspace{3mm} \textbf{if} (($state = \texttt{RF} \vee \texttt{FC})$ $\wedge$ $\neg catchup.$isEmpty) 
\STATE \hspace{6mm} \textbf{return} $catchup$.front
\STATE \hspace{3mm} \textbf{else if} (($state = \texttt{FC})$ $\wedge$ $catchup.$isEmpty) 
\STATE \hspace{6mm} \label{line:toNormal} $state \leftarrow \texttt{N}$
\STATE \hspace{6mm} \textbf{return} $normal$.front
\STATE \hspace{3mm} \textbf{else if} ($state = \texttt{N}$ $\wedge$ $\neg normal.$isEmpty)
\STATE \hspace{6mm} \textbf{return} $normal$.front
\STATE \hspace{3mm} \label{line:emptyFront} \textbf{else} \textbf{return} $\perp$

\vspace{2mm}

\STATE \textbf{pop}
\STATE \hspace{3mm} \textbf{if} (($state = \texttt{RF} \vee \texttt{FC})$ $\wedge$ $\neg catchup.$isEmpty) 
\STATE \hspace{6mm} $catchup$.pop
\STATE \hspace{6mm} \textbf{if} ($state = \texttt{FC}$ $\wedge$ $catchup.$isEmpty)
\STATE \hspace{9mm} $state \leftarrow \texttt{N}$
\STATE \hspace{3mm} \textbf{else if} ($state = \texttt{N}$ $\wedge$ $\neg normal.$isEmpty)
\STATE \hspace{6mm} $normal.pop$

\vspace{2mm} 

\STATE \textbf{suspend}
\STATE \hspace{3mm} lock ($L_q$)
\STATE \hspace{3mm} $state \leftarrow  \texttt{S}$
\STATE \hspace{3mm} \textbf{while} ($\neg normal$.isEmpty)  $normal.$pop
\STATE \hspace{3mm} \textbf{while} ($\neg catchup$.isEmpty) $catchup.$pop
\STATE \hspace{3mm}  delete and clean popped entries
\STATE \hspace{3mm} \textbf{delete} currentBatch
\STATE \hspace{3mm} unlock ($L_q$)

\vspace{2mm}

\STATE \textbf{fetchingCompleted} ($term$)
\STATE \hspace{3mm} \textbf{if} ($state =  \texttt{RF}$ $\wedge current\_term = term$)
\STATE \hspace{6mm} \textbf{if} ($catchup$.isEmpty)  $state \leftarrow \texttt{N}$
\STATE \hspace{6mm} \textbf{else}  
\STATE \hspace{9mm} \label{line:toNormalInFetchingCompleted} $state \leftarrow \texttt{FC}$
\STATE \hspace{9mm} $sendNext (id)$ 

\end{algorithmic}
}
\end{algorithm}

\begin{algorithm} [h]
{
\small
\caption{Target Queue - part 2}
\label{alg:targetQueue2}
\begin{algorithmic} [1]

\STATE \textbf{nextBatch} ($max\_size$)
\STATE \hspace{3mm} lock ($L_q$)
\STATE \hspace{3mm} $currentBatch \leftarrow  \texttt{empty\_batch}$
\STATE \hspace{3mm} $s \leftarrow 0$;
\STATE \hspace{3mm} $f \leftarrow$ front
\STATE \hspace{3mm} \textbf{while} ($s \leq max\_size$ and $f \neq \perp$)
\STATE \hspace{6mm} $currentBatch$.add($f$)
\STATE \hspace{6mm}  pop
\STATE \hspace{6mm} $f \leftarrow$ front
\STATE \hspace{6mm} $s \leftarrow s + 1$
\STATE \hspace{3mm} \textbf{return} $currentBatch$ 
\STATE \hspace{3mm} unlock ($L_q$)

\vspace{2mm}

\STATE \textbf{popBatch} 
\STATE \hspace{3mm} lock ($L_p$)
\STATE \hspace{3mm} $index \leftarrow currentBatch.lastEntry.index$
\STATE \hspace{3mm} popped[$index$] $\leftarrow currentBatch$
\STATE \hspace{3mm} unlock ($L_p$)

\vspace{2mm}

\STATE \textbf{erase} ($index$)
\STATE \hspace{3mm} lock ($L_p$)
\STATE \hspace{3mm} \textbf{delete} $popped[index]$
\STATE \hspace{3mm} $popped$.erase($index$)
\STATE \hspace{3mm} unlock ($L_p$)

\end{algorithmic}
}
\end{algorithm}

\subsection{Dispatching}
\label{sec:dispatcher}
Once the fetcher learns a new log entry, it calls the dispatcher to process the new entry as shown in Algorithm \ref{alg:dispatcher}. A single log entry may require sending multiple messages to different targets. The dispatcher generates the required messages, pushes them to the target queues, and calls $sendNext$ with the ids of the receiving targets. 
Each entry has an index that is assigned by the log service. After successfully consuming an entry, a target writes an acknowledgment with the index of the entry to the stream \footnote{In practice, if consuming entries is expensive (e.g. requires writing to the disk, as in the case of storage shards), the target can buffer and consume a sequence of entries together and send the acknowledgment only for the last entry in the sequence.}. Thus, after sending each entry, the dispatcher calls $readNext$ to read its acknowledgment. The dispatcher avoids pushing an entry $e$ to a queue if the corresponding target has already acknowledged consuming the entry (i.e., $last\_ack[id] >= e.index$). 
Whenever the dispatcher dispatches an entry of the main stream, it updates variable $current\_index$ to the index of the last entry of the main stream that it has dispatched. 


We can write to a gRPC stream only after the previous write has shown up in the completion queue. To guarantee that, we use a $write\_status$ flag for each stream. Upon calling $sendNext$, if the status of the given stream is \texttt{ready}, we obtain the front entry of the corresponding target queue to send. If the front entry is available, we create the next batch of entries and write it to the stream. We include the id of the queue and the operation type (i.e. write) in the tag number. Finally, we change the state of the stream to \texttt{not-ready} to prevent another write to stream before the new write shows up on the completion queue. Similarly, we protect reading from the stream using $read\_status$ flag. We maintain the responses from targets for read operations in $responses$. Both $sendNext$ and $readNext$ must be protected by locks, because they may be called by multiple dispatcher threads and the completion queue processing thread (as we will explain below).

\begin{algorithm} [h]
{
\small
\caption{Dispatcher}
\label{alg:dispatcher}
\begin{algorithmic} [1]

\STATE \textbf{dispatch} (log entry $e$, $isNormal$, $term$)
\STATE \hspace{3mm} lock ($L_d$)
\STATE \hspace{3mm} \textbf{for} (each $id$ in  $e.target\_ids$)
\STATE \hspace{6mm} \label{line:check}\textbf{if} ($last\_acks[id] <  e.index$)
\STATE \hspace{9mm} \label{line:push} $target\_queues[id]$.push($m$, $isNormal$, $term$)
\STATE \hspace{9mm} sendNext($id$)
\STATE \hspace{9mm} readNext($id$)
\STATE \hspace{3mm} \textbf{if} ($isNormal$) $current\_index \leftarrow e.index$
\STATE \hspace{3mm} unlock ($L_d$)

\vspace{2mm}

\STATE \textbf{sendNext} ($id$)
\STATE \hspace{3mm} lock ($L_s$)
\STATE \hspace{3mm} \textbf{if} ($streams[id].write\_status = \texttt{ready})$
\STATE \hspace{9mm} $batch \leftarrow target\_queues[id]$.nextBatch
\STATE \hspace{6mm} \textbf{if} ($batch \neq  \texttt{empty\_batch}$ $\wedge state \neq  \texttt{S})$
\STATE \hspace{9mm} $streams[id].$write($batch$, $\langle id, write \rangle$)
\STATE \hspace{9mm} $streams[id].write\_status \leftarrow \texttt{not-ready}$
\STATE \hspace{3mm} unlock ($L_s$)
\vspace{2mm}

\STATE \textbf{readNext} ($id$)
\STATE \hspace{3mm} lock ($L_r$)
\STATE \hspace{3mm} \textbf{if} ($streams[id].read\_status= \texttt{ready})$
\STATE \hspace{6mm} $streams[id].$read ($responses[id]$, $\langle id, read \rangle$)
\STATE \hspace{6mm} $streams[id].read\_status \leftarrow \texttt{not-ready}$
\STATE \hspace{3mm} unlock ($L_r$)
\end{algorithmic}
}
\end{algorithm}

\subsection{Processing the Completion Queue}
\label{sec:cq}
As gRPC processes the read and write requests, their tags show up on the completion queue. As explained in Section \ref{sec:dispatcher}, we encode the target id and the operation type (i.e., read or write) in a single integer and use it as the tag. We use a separate thread to process the completion queue. We call this thread the completion queue consumer thread. It is shown by a circle in Figure \ref{fig:arch} and its behavior is provided in Algorithm \ref{alg:cq}. It continuously reads the next entry of the completion queue. When the completion queue is empty, its consumer thread gets blocked on $cq.next$ until the next entry is available. 
When a write shows up on the completion queue, it means it is safe to write the next entry. However, showing up a write on the completion queue does not mean that gRPC is done with sending the corresponding message. Thus, we cannot delete any memory allocated to the batch. Instead, we just pop it from the queue, and set $write\_status$ to \texttt{ready} signaling that the stream is ready for writing the next entry. We obtain the id of the right queue via the id that is encoded in the tag. 
When a read shows up on the completion queue, it means gRPC has completed a read operations and the response is ready in $responses$. We access the right entry of the $responses$ via the id that is encoded in the tag. After receiving a response, we update the proper entry of $last\_acks$ with the index of the response. We also delete the memory of the batch associated with the response in the map $popped$ of the target queue, as gRPC does not need this batch anymore. Then, we set $read\_status$ to \texttt{ready} and request gRPC to read the next response for this target.

\begin{algorithm} [h]
{
\small
\caption{Completion queue consumer thread}
\label{alg:cq}
\begin{algorithmic} [1]

\STATE \textbf{while} ($tag$ = $cq$.next) 
\STATE \hspace{3mm} \textbf{switch} ($tag.kind$)

\STATE \hspace{6mm} \textbf{case:} write
\STATE \hspace{9mm} $target\_queues[tag.id]$.popBatch
\STATE \hspace{9mm} $streams[tag.id].write\_status \leftarrow \texttt{ready}$
\STATE \hspace{9mm} sendNext($id$)

\vspace{1mm}

\STATE \hspace{6mm} \textbf{case:} read
\STATE \hspace{9mm} $response \leftarrow response[tag.id]$
\STATE \hspace{9mm} $last\_acks[id] = responses.index$
\STATE \hspace{9mm} $target\_queues[tag.id].$erase($response.index$)
\STATE \hspace{9mm} $streams[tag.id].read\_status \leftarrow \texttt{ready}$
\STATE \hspace{9mm} readNext($id$)
\end{algorithmic}
}
\end{algorithm}

\subsection{Recovery}
\label{sec:recovery}

In this section, we focus on how \name handles faults. We consider two cases. First, we see how \name handles its crash (i.e. how \name restarts). Then, we focus on the case where a target comes back after a crash or disconnection from \name. Finally, we provide an optimization to improve the performance of both recovery cases. The recovery procedures are shown in Algorithm \ref{alg:recovery}.

\subsubsection{\name Recovery}
\label{sec:restart}
Once \name restarts, it first asks targets the highest index of entries each of them has consumed so far and puts responses in $last\_acks$. Then, it sets the minimum of all entries in $last\_acks + 1$ as the $start\_index$ and creates a new fetcher service that starts reading log entries from the $start\_index$. It specifies type normal for this fetcher, as it is the main stream of \name. This will cause this fetcher to set $inNormal$ to \texttt{true} when it calls $dispatch$ function of the dispatcher (see Algorithm \ref{alg:dispatcher}). By starting from the minimum entry of $last\_ack + 1$, we guarantee that no target misses any entry due to a \name crash.

\subsubsection{Target Recovery}
\label{sec:targetRecovery}
Once the health checker finds out that a target is down/disconnected, it sets the status of its queue to suspended. Once the target $i$ is up/connected again, we first increment the term of the target and change the state of the corresponding queue to $\texttt{RF}$. Changing state to $\texttt{RF}$ guarantees that we can push new entries to the queue (see Algorithm \ref{alg:targetQueue1}). 
When \name re-establishes its connection with the target, it asks the target the highest index that it has consumed and puts it in $last\_asks[i]$. If $current\_index$ is greater than $last\_asks[i]$, it means the target has potentially missed some entries. In this situation, we create a new fetcher service that reads entries from $last\_acks[i] + 1$ to the $current\_index$. Note that it is guaranteed that any entry with an index greater than $current\_index$ will be pushed to the queue via the normal stream. Thus, the recovery stream only need to read from $last\_acks[i] + 1$ to  $current\_index$. If $current\_index$ is equal $last\_acks[i]$, it means the target has not missed any entry while it was down. Thus, we just set the state of the queue to $\texttt{N}$. To prevent the procedure of dispatching new entries and updating $current\_index$ from interfereing with the procedure of starting the recovery fetcher, we protect them with the same lock $L_d$ (see Algorithm \ref{alg:dispatcher}). 

\subsection{Improving Recovery Performance}
\label{sec:dummies}
The recovery procedures provided in Section \ref{sec:restart} and \ref{sec:targetRecovery} guarantees that targets do not miss any entries. However, both recovery cases suffer from targets that are long behind the normal stream. Specifically, if a target $i$ does not receive an entry for a long time, its $last\_acks[i]$ remains significantly smaller than the index of the main stream. Now, if \name restarts, according to Algorithm~\ref{alg:recovery}, we have to start the main stream from a very old index determined by $last\_acks[i]$. However, none of the targets needs to start its recovery stream from that index. Similarly, if target $i$ crashes and comes back, we have to start its recovery from an old index while it is not necessary because the target did not have any entries for a while. To solve this issue, we send dummy entries to targets once in a while to advance the index of the last entry received by them.

\begin{algorithm} [h]
{
\small
\caption{Recovery}
\label{alg:recovery}
\begin{algorithmic} [1]
\STATE \textbf{Upon} \name restart
\STATE \hspace{3mm} \textbf{for} (each target i) 
\STATE \hspace{6mm} $last\_ack[i] =$ getLastAck($i$) 
\STATE \hspace{6mm} $start\_index \leftarrow$ min ($last\_acks$) + 1
\STATE \hspace{6mm} new fetcher ($start\_index$, normal)

\vspace{2mm}

\STATE \textbf{Upon} target $i$ restart
\STATE \hspace{3mm} lock ($L_d$)
\STATE \hspace{3mm} $term \leftarrow term + 1$
\STATE \hspace{3mm} $target\_queues[i] \leftarrow$ $\texttt{RF}$
\STATE \hspace{3mm} $streams[i].write\_status \leftarrow$ ready
\STATE \hspace{3mm} $last\_acks[i] \leftarrow$ getLastAck()
\STATE \hspace{3mm} \textbf{if} ($last\_acks[i] < current\_index$)
\STATE \hspace{6mm} $end\_index \leftarrow current\_index$
\STATE \hspace{6mm} new fetcher ($last\_acks[i] + 1$, $end\_index$, recovery)
\STATE \hspace{3mm} \textbf{else}
\STATE \hspace{6mm} $target\_queues[i] \leftarrow$ $\texttt{N}$
\STATE \hspace{3mm} unlock ($L_d$)
\vspace{2mm}

\STATE \textbf{Upon} dispatching every $E$ entries 
\STATE \hspace{3mm} \textbf{for} (target\_queue $i$ : $target\_queues$) 
\STATE \hspace{6mm} \textbf{if} $last\_acks[i] < current\_index - E$ 
\STATE \hspace{9mm} $dummy \leftarrow $ $\texttt{empty\_message}$
\STATE \hspace{9mm} $dummy.index \leftarrow current\_index$
\STATE \hspace{9mm} $target\_queues[i].$push($dummy$) 

\end{algorithmic}
}
\end{algorithm}
 \section{Correctness}
 \label{sec:correctness}
 
 In this section, we focus on the correctness of our design. We first explain what fault-tolerance requirements we want our \name to satisfy. Then, we model check the correctness of our design using TLA+ \cite{tla}. 
 
 \subsection{Fault-tolerance}
 \label{sec:faultTolerance}
 The purpose of fault-tolerance is to continue satisfying the desired specification even in the presence of faults that perturb the program from its normal behavior. In case of our \name, we consider the in-order exactly-once delivery as the desired specification. This specification consists of two parts: 1) each log entries is eventually delivered to its targets, 2) no entry is delivered to a target more than one time or out of the order. The first property is a \textit{liveness} property while the second one is a \textit{safety} property \cite{liveness}. The failures of targets or \name are considered as faults for the overall design. Note that the first property can only be satisfied if the faults stop occurring (or at least happen with large enough intervals). In other words, if components keep failing, it is impossible to have an algorithm that satisfies the first property. Thus, we want our \name to satisfy the following two properties: 

\begin{enumerate}
    \item After any execution with any number of faults, if faults stop occurring, each entry (added before or after faults stop) is eventually delivered to its targets. 
    \item In the presence of faults, no entry is delivered to a target more than once or out of the order. 
\end{enumerate}

The first property is a \textit{non-masking} fault-tolerance property that requires that the program must return to its desired behavior after being perturbed by faults. The second one is a \textit{failsafe} fault-tolerance property that requires that the program cannot violate its safety property in presence of faults. The combination of both is \textit{masking} fault-tolerance \cite{masking}. Thus, we want our \name to be masking fault-tolerant to the failures of targets or \name for the in-order exactly-once delivery specification. 
 
 \subsection{Model Checking with TLA+}
 \label{sec:tla}
 We can directly model a program in TLA+ \cite{tla}. However, a more convenient way is using PlusCal language that is more similar to programming languages than TLA+. The PlusCal code can then be translated to a TLA+ specification using TLA+ toolbox. Here, we do not want to explain how to describe a concurrent program in PlusCal language (the reader may refer to \cite{tla}). However, we cover those parts that are necessary to understand the PlusCal code presented in this section. Specifying a program for model checking requires some art to avoid creating models with too large state spaces. Specifically, if we model all details of a program, the model may become too large causing the model checking process to take too much time. One approach to reduce the burden of model checking is to avoid modeling parts that are identical. For our \name, for instance, the behavior of each target is the same. Thus, it is enough to model a system with only one target. Another approach is to abstracting away parts that are not necessary to check based on our assumptions. For example, we abstract away the details of gRPC and network communications. Also, we cannot model the program for an infinite number of log entries or any number of failures. Thus, we model check the program for sending a limited number of log entries and failures. We denote this parameters by $\texttt{NMESSAGES}$ and $\texttt{NFAILURES}$.  
 
\begin{figure}
\begin{center}
\includegraphics[width=\x\columnwidth]{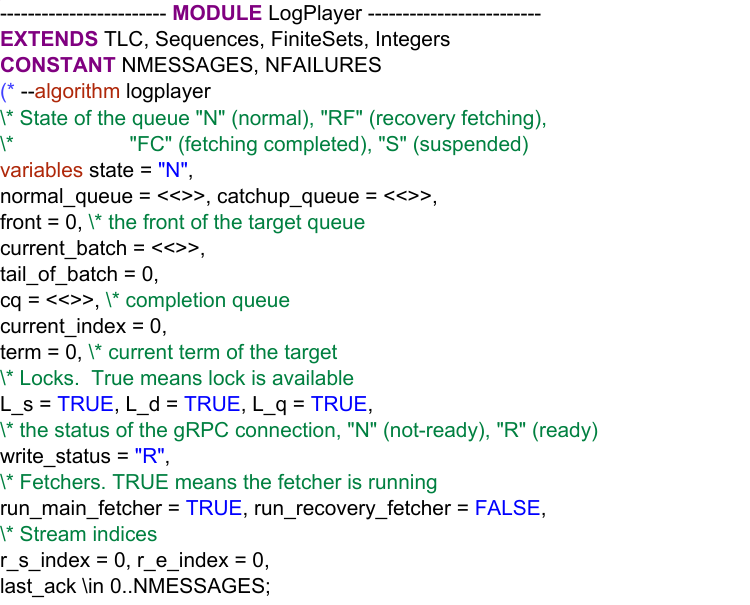}
\caption{PlusCal code variables definitions}
\label{fig:plusca1:variables}
\end{center}
\end{figure}

\begin{figure}
\begin{center}
\includegraphics[width=\x\columnwidth]{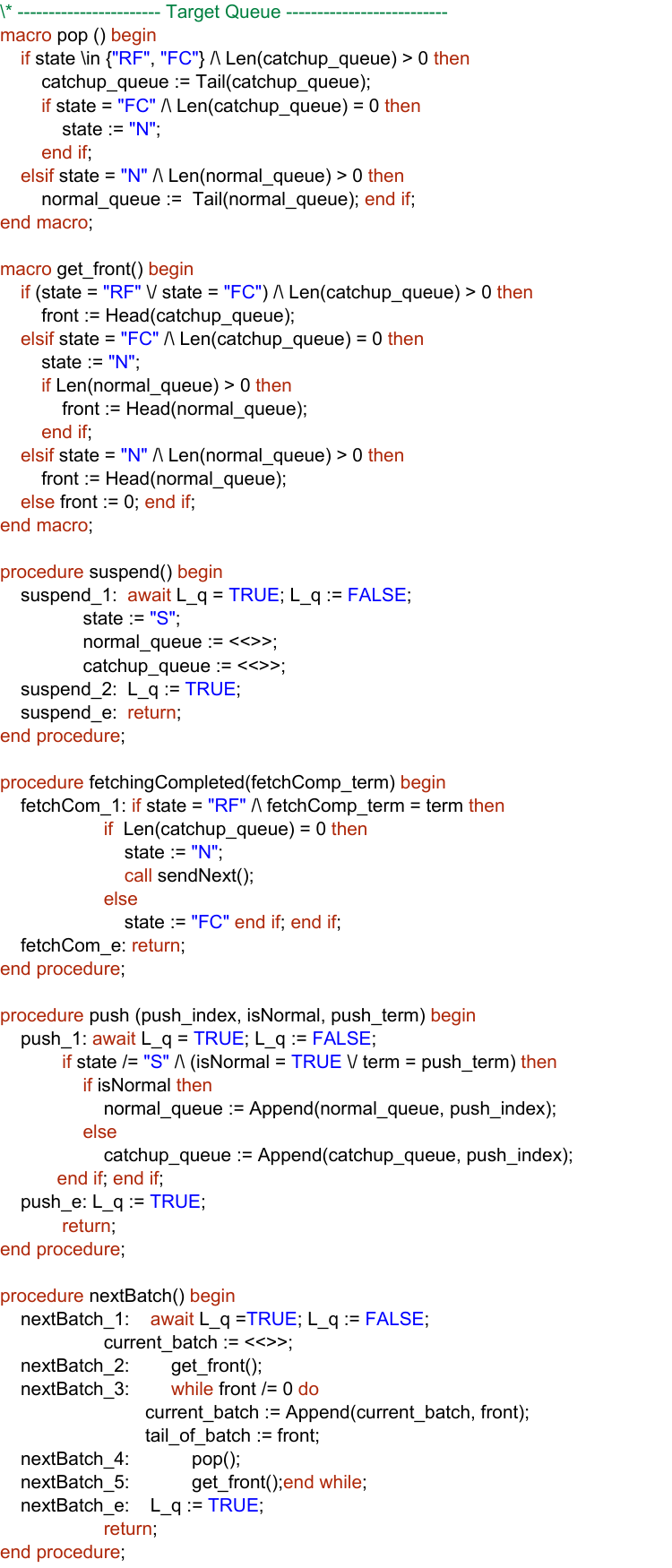}
\caption{PlusCal code of the target queue}
\label{fig:pluscal:targetqueue}
\end{center}
\end{figure}

\begin{figure} [h]
\begin{center}
\includegraphics[width=\x\columnwidth]{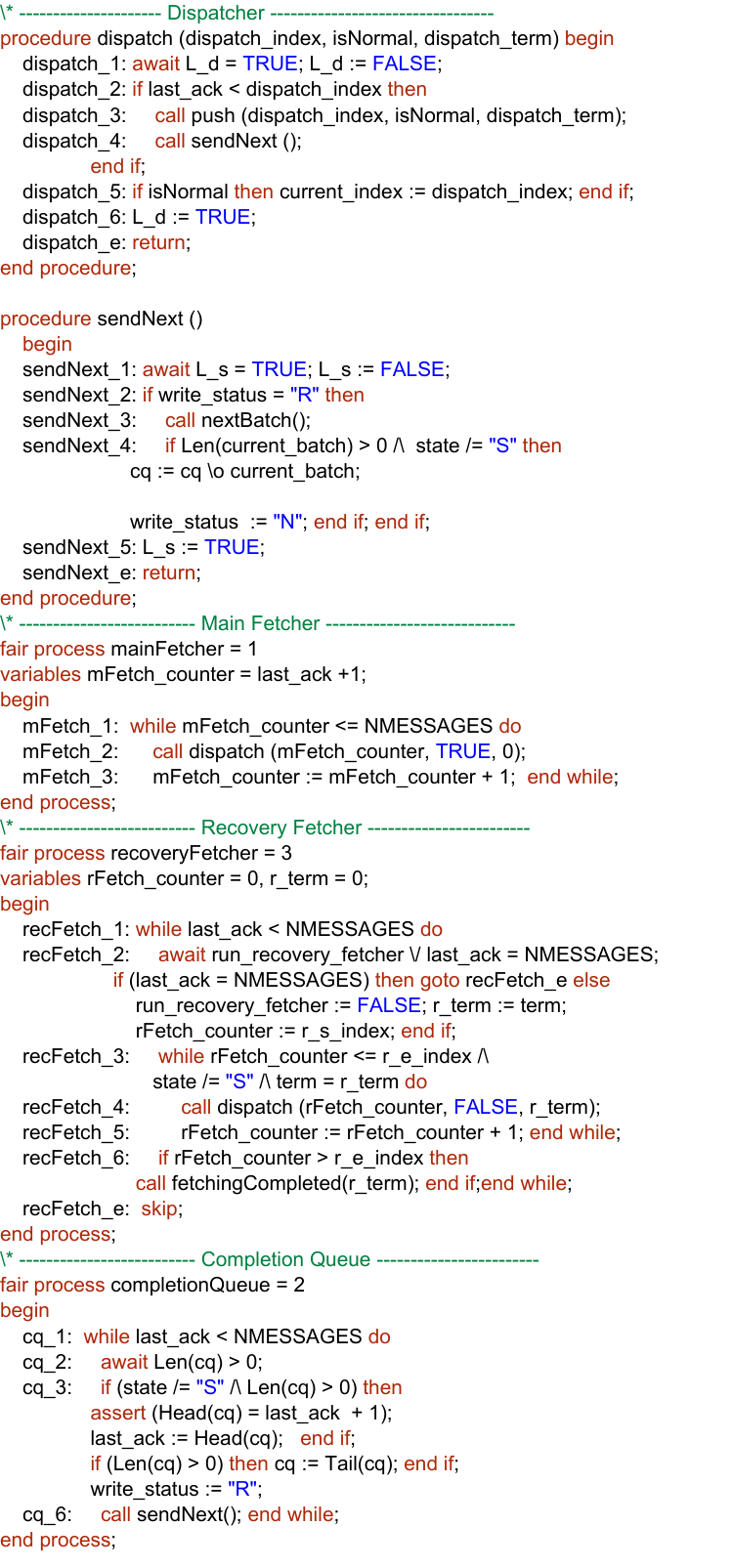}
\caption{PlusCal code of dispatcher, fetchers, and completion queue}
\label{fig:pluscal:dispatcher}
\end{center}
\end{figure}

 Figure \ref{fig:plusca1:variables} shows the variables and their initial values. Each state of the program corresponds to some assignment to its variables. The variables $normal\_queue$, $catchup\_queue$, and $current\_batch$ are defined as tuples that can be used to simulate queues. The initial value of $last\_ack$ can be any number in the range of $[0, \texttt{NMESSAGES}]$. A value $x$ greater than 0 for $last\_ack$ models the case where \name has restarted and learned that the last entry received by the target is $x$ (see Algorithm \ref{alg:recovery}).
 %
 Figure \ref{fig:pluscal:targetqueue} describes the logic of our target queue in PlusCal. Using labels we can specify the level of atomicity of our operations.  All operations associated with a single label are executed atomically. Also, the whole body of a macro is executed atomically. Thus, there is no label inside macros. The more labels we have, the larger the state space is which consequently results in the longer model checking time. Thus, to keep model checking time low, we have to avoid adding labels as much as possible without eliminating important concurrency aspects that we are interested to check. Therefore, for $pop$ and $get\_front$ operations we use macro, as we are sure that they are only used in $nextBatch$ function that is protected by $L_q$ lock (see Algorithm \ref{alg:targetQueue2}). The PlusCal code provided in Figure \ref{fig:pluscal:targetqueue} is easy to understand, as it is very similar to the pseudocode provided in Algorithms \ref{alg:targetQueue1} and \ref{alg:targetQueue2}. 
The PlusCal description of the dispatcher is provided in Figure \ref{fig:pluscal:dispatcher}. As mentioned above, we do not model the gRPC details. Thus, when the $sendNext$ writes to the stream, we model it as if it appears immediately on the completion queue.  

All PlusCal codes we explained so far are part of either macros or procedures. Now, we explain our \textit{processes}. Processes are the units that can be used to model processes and threads of a concurrent program. This is where PlusCal makes modeling the non-determinism very easy. Specifically, the operations of different processes can interleave in any order, as it may happen for a concurrent program in practice. The model checker considers any possible interleaving and checks the correctness of our program in all paths. We have four processes. All processes are marked with $fair$ keyword which means our processes never stop executing when they can run an operation. The first process is our $mainFetccher$ process that models the fetcher of the main stream. This process simply starts from index $last\_ack + 1$ and calls $dispatch$ for each index up to $\texttt{NMESSAGES}$. The main fetcher runs only one time. Thus, once it dispatched all messages, it finishes. In addition to the main fetcher, we have $recoveryFetcher$ process that models the recovery fetcher. Note that since we may run the recovery fetcher multiple times, the whole process is inside a while loop that finishes once the program is done with sending the number of intended messages. At the beginning of this loop, the process waits for the recovery signal. This way, we model calling the recovery fetcher once the target comes back. 

The next process is the completion queue consumer thread (see Algorithm \ref{alg:cq}). We do not model read operations and network communications. Thus, when an index shows up in front of the completion queue, we check if it is the correct index that the target must have received. We check this using the $assert$ instruction. Note that any duplicate or out of the order delivery makes this assert fail. Thus, here we check the safety part of our specification.

\begin{figure} 
\begin{center}
\includegraphics[width=\x\columnwidth]{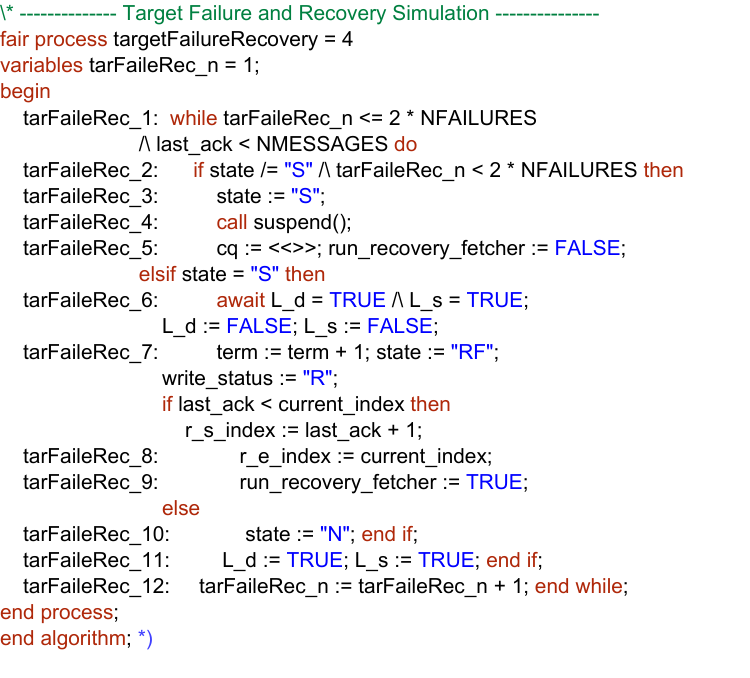}
\caption{PlusCal code of failure and recovery}
\label{fig:pluscal:failure}
\end{center}
\end{figure}

Figure \ref{fig:pluscal:failure} models the failure and recovery of the target. The number of failures by this process is given by the model parameter $\texttt{NFAILURES}$. The conditions of $if$ and $elsif$ are specified such that each failure has a recovery. When a target fails, any message destined at the target will show up on the completion queue as failure. We model this by cleaning the $cq$ upon failure. Also, we change the state of the queue to \texttt{S} and call $suspend$ routine. The rest of the process is the same as the pseudocode provided in Algorithm~\ref{alg:recovery}.  

We check the liveness part of our specification by specifying the following temporal logic property for TLA+   that requires that the target must eventually receive all entries: $\lozenge (last\_ack = \texttt{NMESSAGES})$.
When we model check our TLA+ model, the TLA+ toolbox does not find any deadlock or assertion/property failure. This means that our design guarantees in-order exactly-once delivery of all messages in the presence of failures of the target or \name. This was not the case for the first draft of our algorithms. TLA+ \cite{tla} helped us to find bugs in our design. For example, in the first version of our algorithms, we did not have the term for the targets. We added this variable to satisfy delivery requirements after finding their violation via TLA+. Some deadlocks due to concurrency bugs were also detected by TLA+ that are fixed in algorithms presented in this paper. 

\section{Configuring LogStore and \name}
\label{sec:wal}
\begin{figure}
\begin{center}
\includegraphics[width=\archScale\columnwidth]{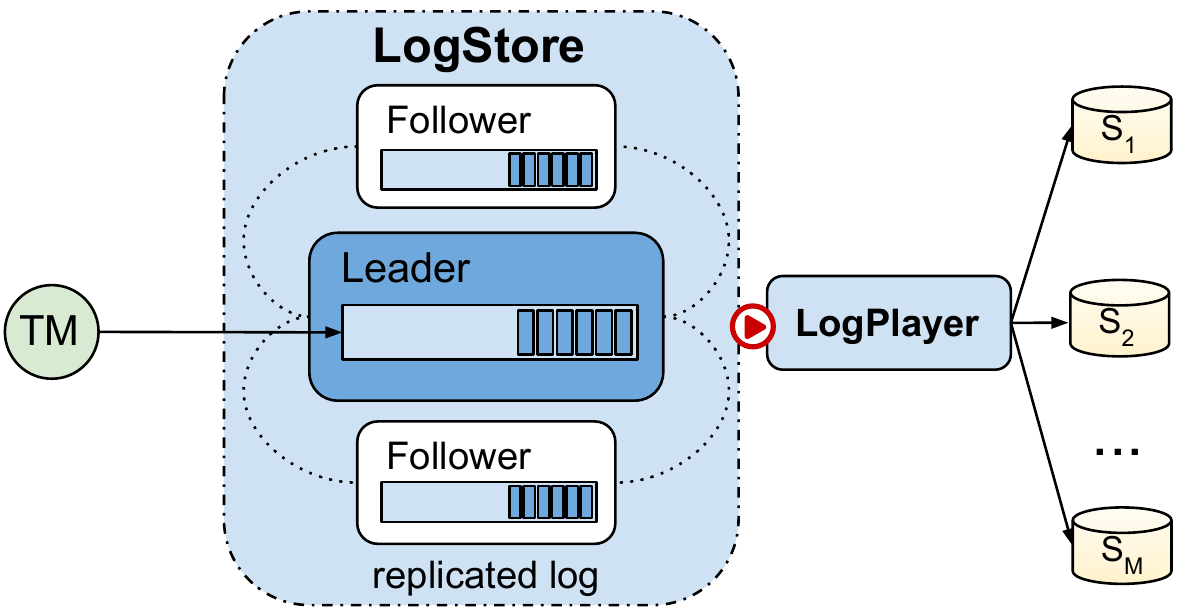}
\caption{Configuring LogStore and \name}
\label{fig:single}
\end{center}
\end{figure}

As explained in Section \ref{sec:intro}, the purpose of a WAL is to guarantee the atomicity and durability of transactions. In addition to these requirements, we want our WAL to be highly available, i.e., our WAL must be able to quickly resume its normal operation in case of a failure. Figure \ref{fig:single} shows how \name can be connected to the LogStore \cite{logstore} and storage shards.
Our desired requirements of atomicity, durability, and high-availability are satisfied by the following features: 1) Writes to LogStore are durable, 2) LogStore is highly-available by replicating the log on several machines, 3) \name guarantees in-order exactly-once delivery even in presence of failures of targets or itself, and 4) \name can easily restart or be replaced by a new instance in case of a failure.

LogStore maintains a replicated log using raft consensus protocol \cite{raft}. This way, it guarantees that the log data is replicated on several machines and all replicas have the same order of log entries. LogStore provides high availability by changing the leader in case of leader failure. The raft protocol guarantees that the new leader has always the most recent log entries \cite{raft}. Thus, once we successfully write to the leader, we can rest assured that our write is durable. Although to append to the WAL, the transaction manger has to send the entry to the leader LogStore server, \name can use any of the LogStore replicas to read the WAL entries. 

The durability of the LogStore and exactly-once delivery of \name guarantee that once we successfully append a new transaction to the LogStore, the mutations required by the transaction will be applied to the storage shards exactly one time, even in presence of failure of \name or storage shards. However, there is still a possibility of duplication in our WAL due to the failure of LogStore or the transaction manager. For example, suppose the transaction manager writes a transaction to the LogStore, but it crashes before receiving the acknowledgment from the LogStore. In this case, the transaction manager may want to try again which will result in duplication, i.e., transaction will be appended and subsequently executed two times. Avoiding duplicates due to LogStore/transaction manager failures is out of the scope of this paper, but one approach is to make updates to the LogStore \textit{idempotent} \cite{book} as follows: we can assign a sequence number to each write by the transaction manager to the LogStore. The LogStore can, then, ignore any message with a duplicate sequence number from a transaction manager. 

\section{Related Work}
\label{sec:related}

Apache Kafka \cite{kafka} and Apache Flink \cite{flink} are two open-source streaming platforms that have recently started supporting exactly-once semantics. In this section, we briefly review these systems and compare them with our \name. 

\subsection{Kafka Transactions and Exactly-once Semantics}
\label{sec:kafka}
Apache Kafka \cite{kafka} is a popular streaming platform widely used for designing high-throughput streaming systems. Kafka has recently started providing transactions that let us atomically write to multiple Kafka partitions. These partitions can belong to different topics. For writes in the context of a transaction, Kafka guarantees either all or none of them will be eventually visible to the consumer. Kafka achieves this by running the Two-Phase Commit (2PC) protocol over partitions involved in a transaction~\cite{kafka}. 
Kafka transactions are the key for providing the atomic \textit{read-process-write} cycles. Typically, a stream processing application consumes some records from input streams, process them, and writes some records to the output streams. To guarantee that each record from the input stream is consumed and processed one time, and resulted records are delivered to the output streams exactly one time, we need to make sure that the whole read-process-write cycle is done atomically, i.e., the application should not reach a state where a record is marked as consumed, but resulted records are not written to the output streams or vice versa. Kafka marks a record as consumed by writing to a special internal topic called \textit{offsets} topic. Using the atomicity of Kafka transactions, we can guarantee that marking a record as consumed and appending produced records to the target partitions (i.e., the read-process-produce cycle) happen atomically \cite{kafkaBlog}. 

\begin{figure} 
\begin{center}
\includegraphics[width=\columnwidth]{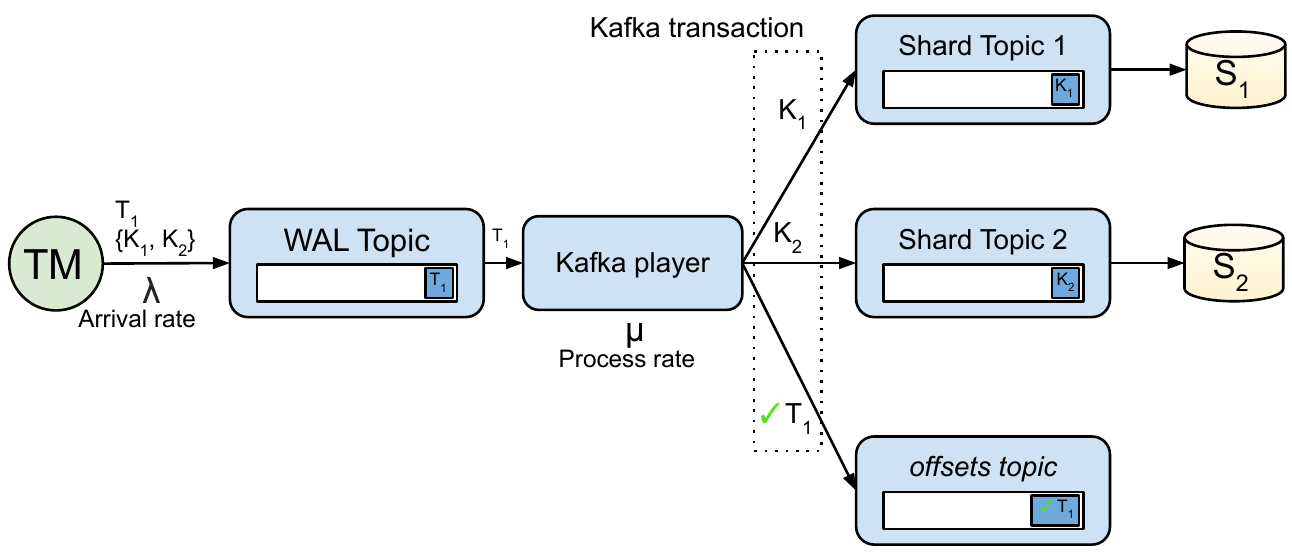}
\caption{WAL with exactly-once guarantee using Kafka transactions.}
\label{fig:kafkaWAL}
\end{center}
\end{figure}

Figure \ref{fig:kafkaWAL} shows how we can design a WAL system using Kafka. We use a Kafka topic as the \textit{WAL topic}. This topic should have only one partition, as the order of transactions should be kept on the WAL, and Kafka does not maintain the global ordering over multiple partitions. In addition to the WAL topic, we have a topic corresponding to each storage shard. We refer to these topics as \textit{shard topics}. Now, to apply mutations recorded on the WAL topic to the storage shards, we have an application that reads records from the WAL topic, produces records containing partial mutations for different shards, and writes them to the shard topics. We call this application \textit{Kafka player}. By running the read-process-write cycle of each WAL entry in the context of a single Kafka transaction, we guarantees that mutations are appended to the shard topics exactly one time. Now, we need to guarantee that storage shards also apply the mutations exactly one time. That can be achieved by atomically storing both the offsets and mutations of entries on the storage shards. Now, when a storage shard recovers from a crash, it continues polling entries starting from its last stored offset. This design, however, requires all shard topics to be durable. Otherwise, if a broker crashes after it received shard mutations and before the shard polled them, then the mutations are missed. Thus, in this design, we have to pay the cost of durability (i.e. distributed consensus and/or disk delay) two times; one time for the WAL topic and one time for shard topics. On the other hand, \name does not need to make the produced partial mutations durable, and at the same time, it can recover from crashes without violating exactly-once semantics. 

The design in Figure \ref{fig:kafkaWAL} has a bigger problem that makes it impractical to be used, assuming all topics are run on the same Kafka cluster and all have the same replication factor. Specifically, the rate of processing WAL entries ($\mu$) is expected to be lower than their arrival rate ($\lambda$), because for appending to WAL topic, the transaction manager needs to wait only for the replication, but the Kafka player needs to also wait for 2PC in addition to the replication. This will cause the system shown in Figure \ref{fig:kafkaWAL} to be unstable, i.e., the delay of applying mutations is expected to grow continuously. In Section~\ref{sec:comparisonWithKafka}, we will see how this issue causes very large delays (e.g. more than 10 seconds) after processing just one thousand transactions.

We can solve problems of the design in Figure \ref{fig:kafkaWAL}, by making the transaction manager bypass the WAL topic and Kafka player, and directly write to the shard topics using Kafka transactions. This change basically creates an implicit back-pressure to the transaction manager and solves the instability issue. However, it requires more work by the transaction manager, as it needs to write to multiple topics and run the 2PC which results in higher transaction response time. In Section \ref{sec:results}, we show how our \name achieve better performance compared with this design.

\subsection{Flink Exactly-once Semantics}
\label{sec:flink}
Apache Flink \cite{flink} is a popular stream processing platform. Flink can connect to different storage and streaming systems including Kafka. The fault-tolerance of Flink relies on its checkpointing mechanism which is based on Chandy-Lamport consistent snapshot algorithm~\cite{chandy}. 
Upon a restart, Flink loads the last checkpoint including the states of all operators and the indexes of the input streams, and then continues from there. To avoid duplicate deliveries, Flink requires external sinks to be transactional like Kafka streams. Specifically, sink should provide a way for Flink to write to them in a context of a transaction and to be able to commit or abort the transaction. Using transactional sinks, Flink can avoid delivering duplicate records, by not committing any write before a checkpoint is completed. When we have multiple sinks, Flink must make sure that writes to different sinks all commit or abort together. For this case, Flink relies on 2PC protocol. Since to have Flink end-to-end exactly-once semantics we have to use transactional sink like Kafka, the design of a WAL with Flink would be identical to the one shown in Figure \ref{fig:kafkaWAL}, except we should use Flink instead of the Kafka player. Note that since Flink delays committing writes until checkpointing is completed, the delay of applying mutations is affected by the checkpointing interval in addition to the other delays. Thus, the delay of Flink is expected be even higher than WAL shown in Figure \ref{fig:kafkaWAL} using pure Kafka. 


\section{Experimental Results}
\label{sec:results}

In this section, we evaluate the performance of a WAL system consisting of LogStore and \name. We have developed both of these components in C++. LogStore relies on NuRaft \cite{nuraft} for the log replication. We also compare the performance of our system with the performance of Kafka for providing exactly-once delivery.
Next, we first explain our measures of interest. Then, we explain the experimental setup. Section \ref{sec:numOfTargets} and \ref{sec:txSize} provide experimental results only for our WAL system. Comparison with Kafka is provided in Section \ref{sec:comparisonWithKafka}. 
\subsection{Measures}
\label{sec:measures}
We timestamp each transaction at the following time points: 

\begin{itemize}
    \item Commit time: right before transaction manager sends the append request to the WAL 
    \item Dispatch time: right before \name processes a new log entry.
    \item Apply time: right after a target receives a new transaction to apply. 
\end{itemize}

After a transaction is appended to the WAL, it takes some time for it to be applied to the actual data on the storage shards. We define this time as the \textit{apply delay} and calculate it by subtracting commit time from the apply time.
The time between dispatch to apply is mainly dominated by the delay of processing and delivering by \name. Thus, we call it \textit{\name delay} and calculate it by subtracting dispatch time from the apply time. 

\subsection{Experimental Setup}
\label{sec:setup}

 For both LogStore and Kafka, we set the replication factor to three, i.e., each WAL entry will be copied to three machines by LogStore/Kafka. As we will see, since \name delays are very low i.e., less than one millisecond, we need to compute the delays very precisely. To eliminate any error in our measurements caused by the clock skew between different virtual machines, we run transaction manager, \name, targets, and one of the LogStore/Kafka replicas in one virtual machine, and run other LogStore/Kafka replicas in different virtual machines. In addition to the clock skew, this configuration also removes the network delays from our measurements which is desired, as we want to focus on processing delays rather than network delays that are out of the control of the algorithms. 
 %
 Since the WAL needs the highest level of durability, we configure both LogStore and Kafka for aggressive persistence to the disk. For Kafka, we do that by setting \texttt{flush.messages} to 1 in the topic configuration.  We use virtual machines with 8 vCPUs, Intel Core Processor (Haswell) 2.0 GHz, 16 GB RAM, and 80 GB virtual disk storage. 

\subsection{The Effect of Number of Targets}
\label{sec:numOfTargets}

\begin{figure*}
  \centering
  \subfigure[\name Delay vs. Target\# \label{fig:logplayerDelayTarget}]{\includegraphics[width=\diagramScale\columnwidth]{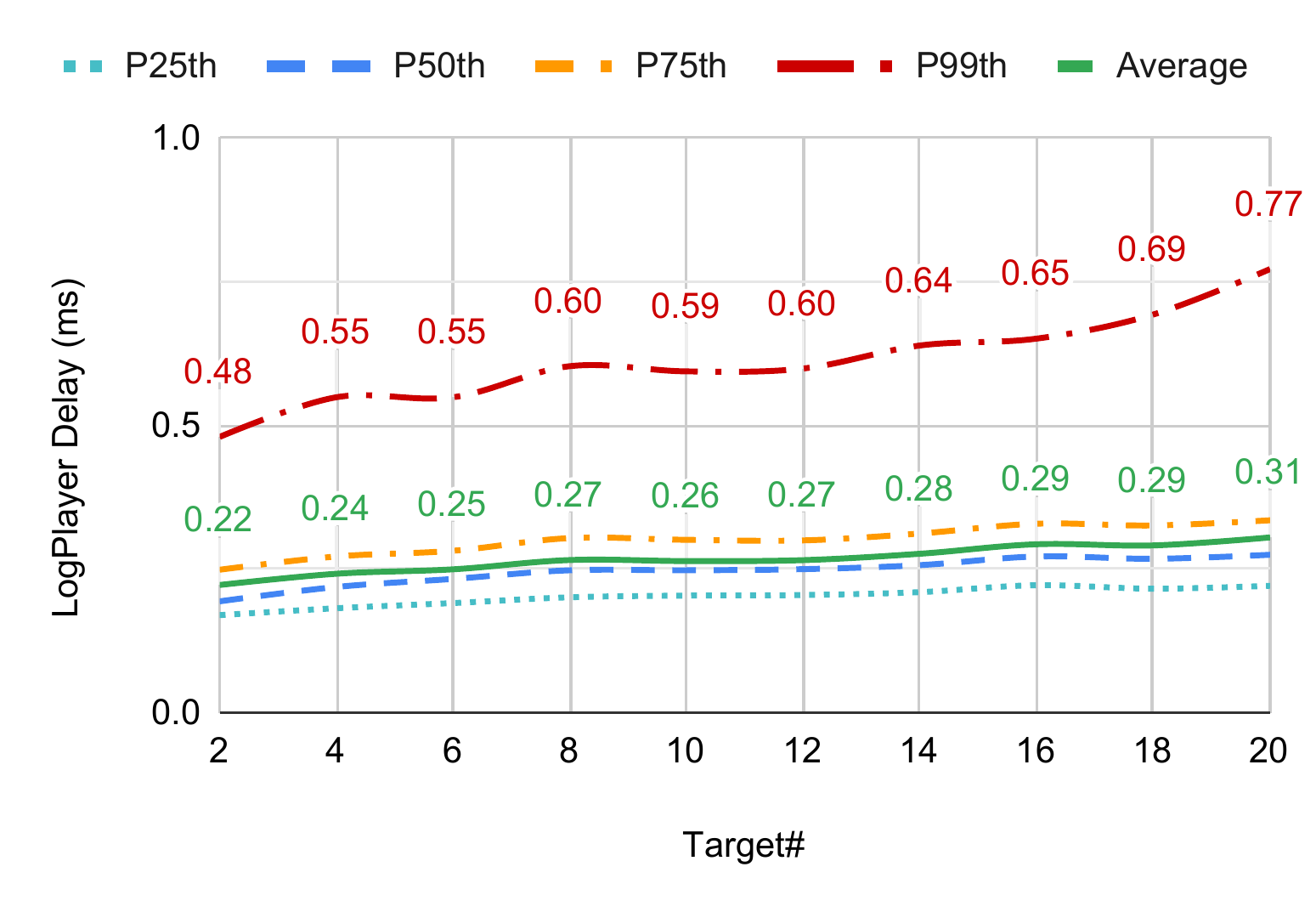}}
  \subfigure[Apply Delay vs. Target\# \label{fig:applyDelayTarget}]{\includegraphics[width=\diagramScale\columnwidth]{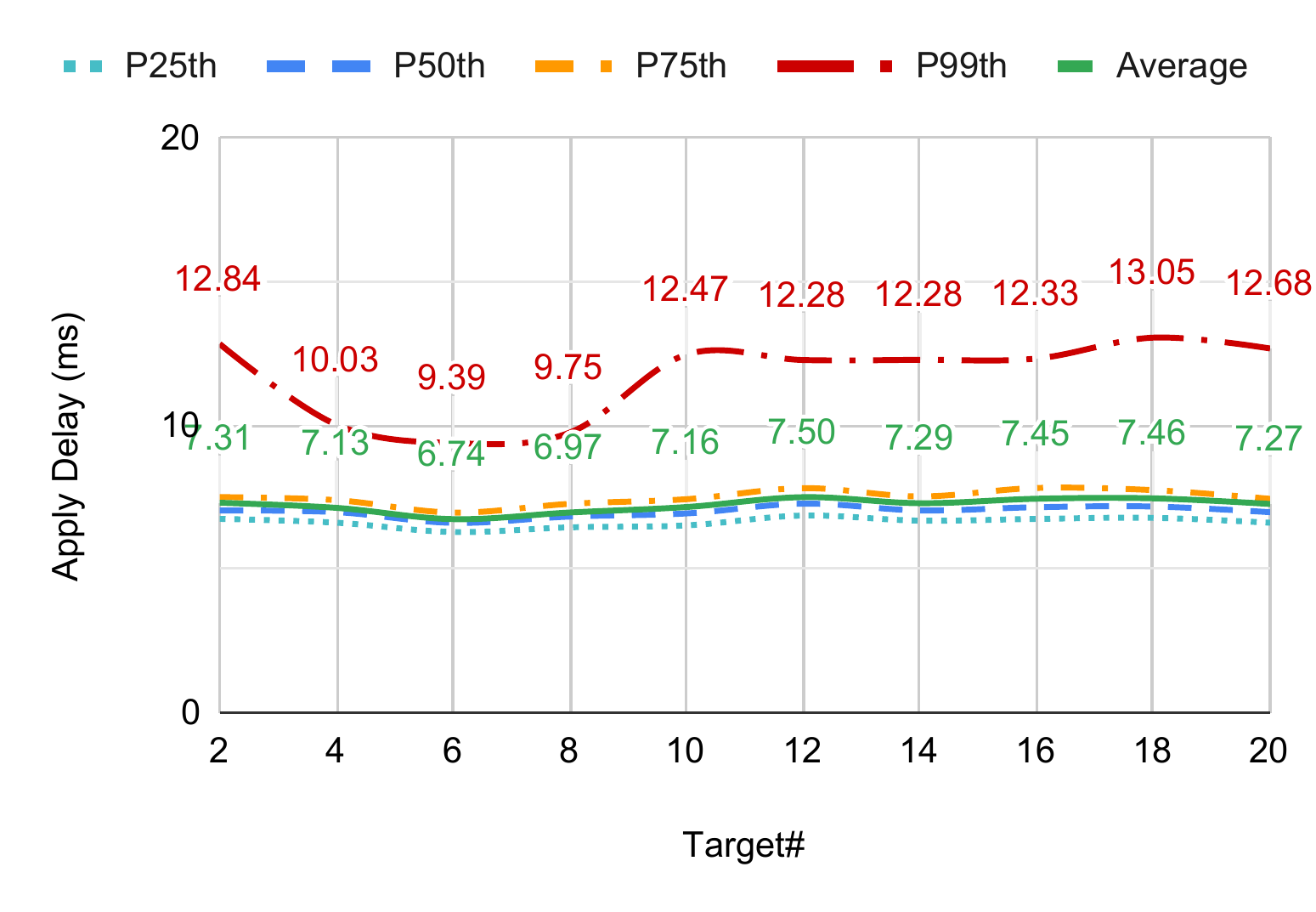}}
 \caption{The effect of number of targets on the \name and apply delay}
\end{figure*}

We first see how our WAL system performs for different numbers of targets. We measure the delays for 2 to 20 targets. 
Figure \ref{fig:logplayerDelayTarget} shows how the average and different percentiles of \name delay change, as we increase the number of targets for transactions of size 10~KB (consisting of 10 key-values each with the total size of 1~KB). The \name delay increases, as we increase the number of targets, because \name has more job for handling more targets. However, the \name delays are very low; the average and median delays remain less than 0.32~ms, and the 99th percentile remains less than 0.78~ms, for all cases.

Figure \ref{fig:applyDelayTarget} shows how the apply delay is affected by the number of targets. The apply delay is completely dominated by the LogStore delay which is resulted by Raft consensus delay and the delay of the persistence to the disk, i.e., the \texttt{fsync} delay. These delays are unaffected by the number of targets. Thus, since sub-millisecond \name delay is negligible compared with LogStore delay, the small increase in the \name delay does not affect the overall apply delay. Thus, we believe the changes to the apply delay as we change the number of targets are due to experimental errors, especially for the 99th percentile that is more sensitive to environmental changes (e.g. variations of the \texttt{fsync} delay).

\subsection{The Effect of Transaction Size}
\label{sec:txSize}

\begin{figure*}
  \centering
  \subfigure[\name Delay vs. Transaction Size \label{fig:logplayerDelaySize}]{\includegraphics[width=\diagramScale\columnwidth]{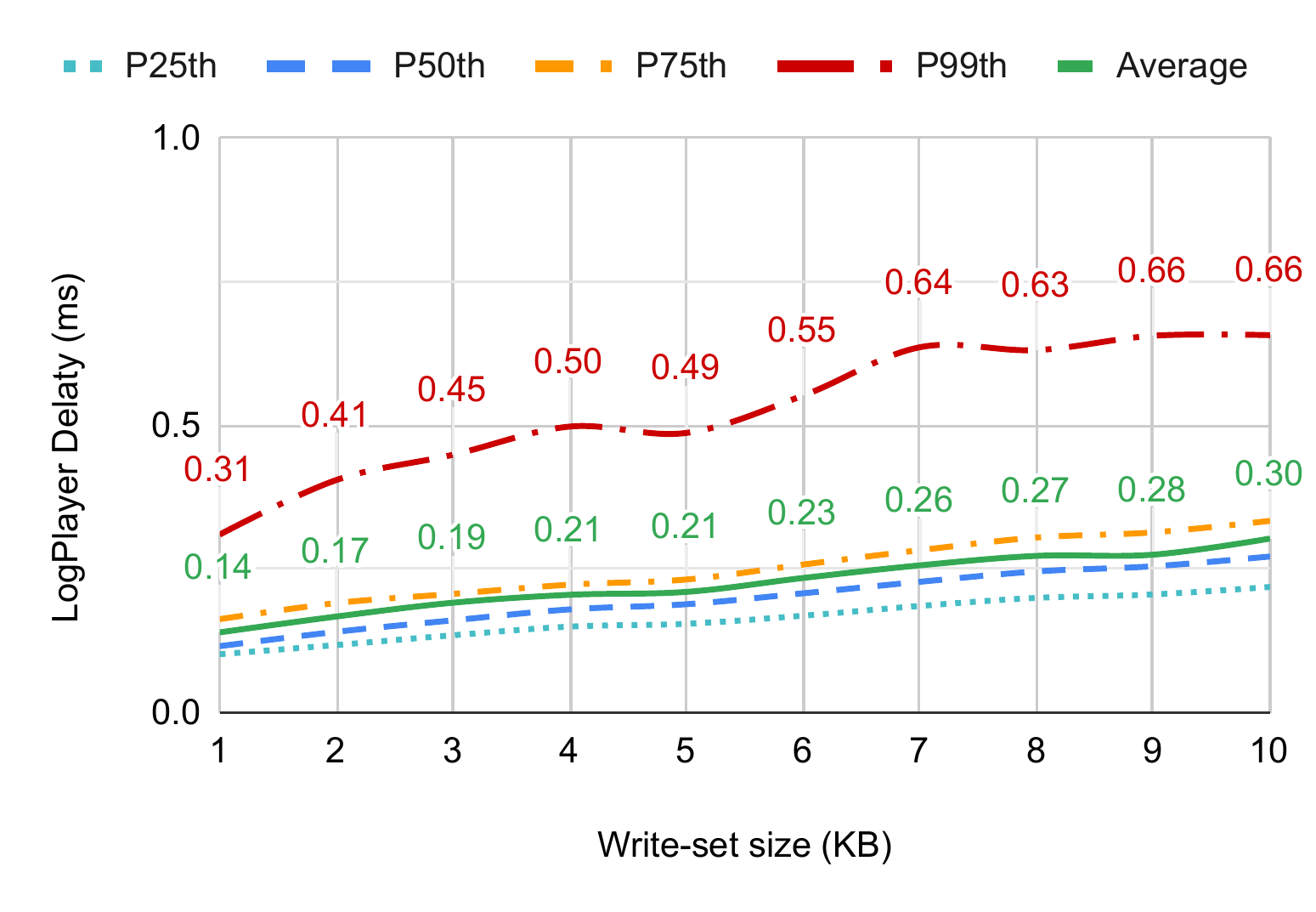}}
  \subfigure[Apply Delay vs. Transaction Size \label{fig:applyDelaySize}]{\includegraphics[width=\diagramScale\columnwidth]{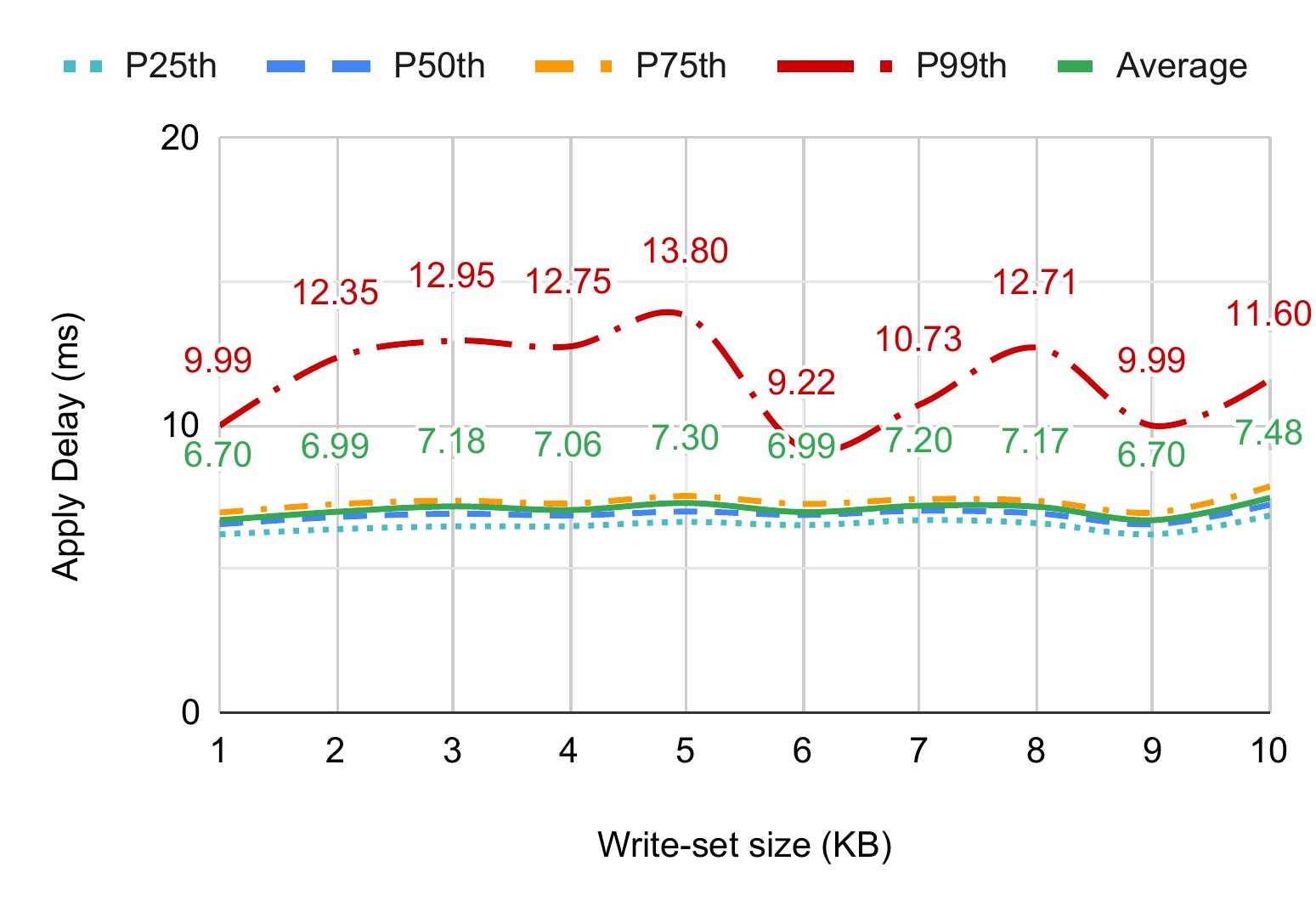}}
\caption{The effect of transaction size on \name and apply delay}
\end{figure*}

Now, we see how the transaction size affects performance. We change the size of transactions from 1~KB to 10~KB for a system with 20 targets. Note that we consider transactions with various numbers of key-values each with 1~KB payload. Figure \ref{fig:logplayerDelaySize} shows how the average and different percentiles of \name delay change, as we increase the size of the transactions. As expected, \name delay increases as we increase the size of the transactions, because \name needs to parse, process, and deliver larger entries. However, again, the \name delay is very low; the average and median remain less than 0.31 ms, and the 99th percentile remains less than 0.67 ms, for all transaction sizes.  
Figure \ref{fig:applyDelaySize} shows the apply delay for different transaction sizes. Although unlike the number of targets, the transaction size is expected to affect the persistence delay, since the persistence delay is largely dominated by the \texttt{fsync} delay, the increase in transaction size shows only a slight increase in the apply delay, e.g. 12\% increase for 10~KB transaction size compared with 1~KB.

\subsection{Comparison with Kafka}
\label{sec:comparisonWithKafka}

\begin{figure} [h]
\centering
 \includegraphics[width=\diagramScale\columnwidth]{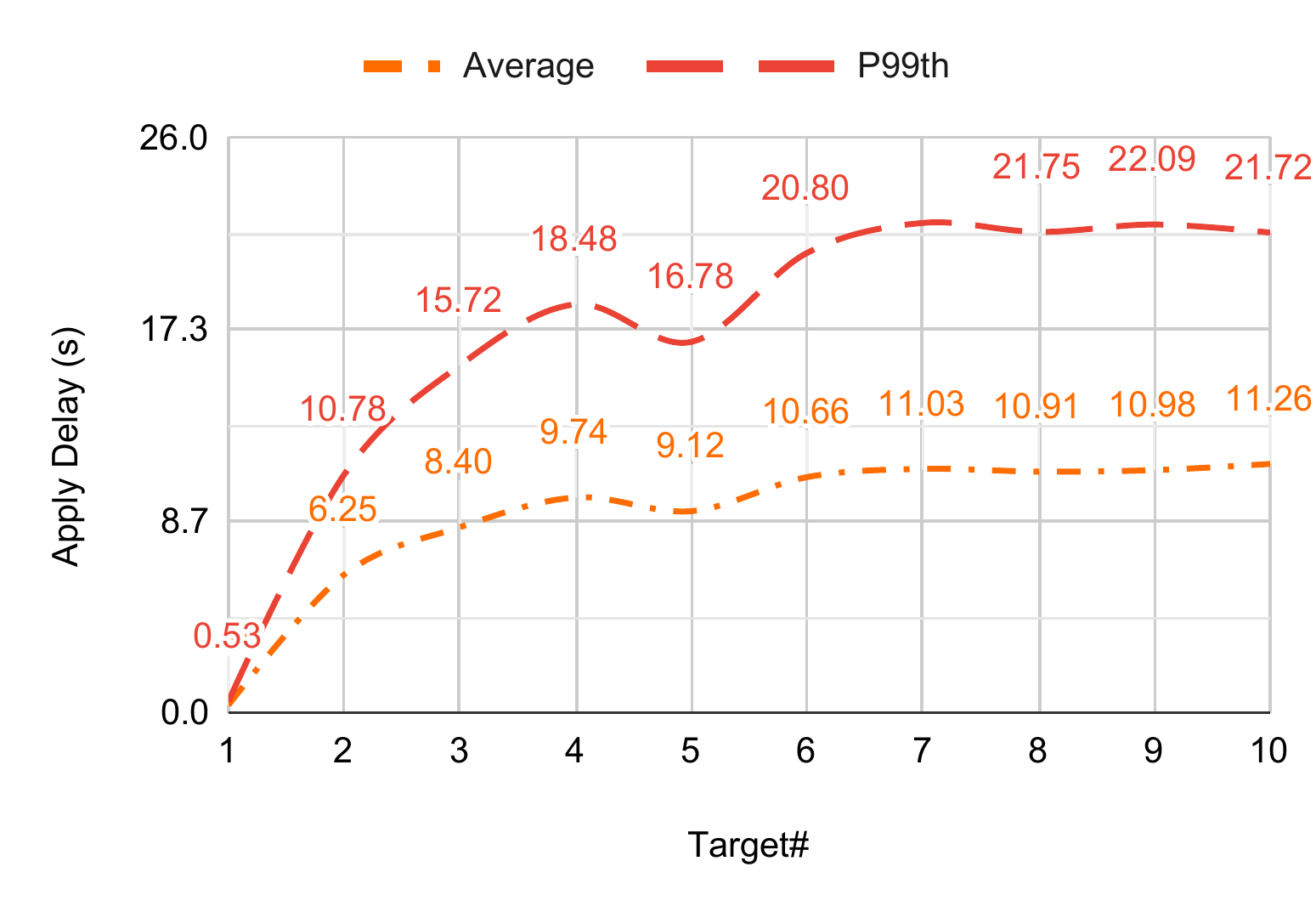}
 \caption{Large apply delay by the design shown in Figure \ref{fig:kafkaWAL} with two levels of Kafka topics}
 \label{fig:applyDelay_target_2level}
\end{figure}

In this section, we compare the performance of a WAL system consisting of LogStore and \name with that of the WAL design with Kafka explained in Section \ref{sec:related}.
First, we want to validate our anticipation of the instability of the design provided in Figure \ref{fig:kafkaWAL}. As explained in Section \ref{sec:related}, the design with 2 levels of Kafka topics is expected to be unstable, as the arrival rate of log entries is expected to be higher than their process rate by the Kafka player. Figure~\ref{fig:applyDelay_target_2level} shows how the average and 99th percentile of apply delay for 1000 transactions changes by increasing the number of targets for transactions with 10 keys. As it is shown in Figure \ref{fig:applyDelay_target_2level}, due to WAL backlog, the average and 99th percentile of apply delay increase to 11.26 and 21.72 seconds, respectively. Note that the diagrams shown in Figure \ref{fig:applyDelay_target_2level} flatten, as we ran the experiments for 1000 transactions. However, if we keep adding entries to the system, the apply delay will grow continuously. This experiment confirms that having two levels of Kafka topics without any back-pressure to the transaction manager is not practical for designing our WAL. Thus, in the rest of the experiments in this section, we use the design where the transaction manager directly appends to the shard topics by bypassing the WAL topic and Kafka player.

\begin{figure*}
  \centering
  \subfigure[Apply Delay vs. Target\# \label{fig:applyDelay_target_kafka_nonT}]{\includegraphics[width=\diagramScale\columnwidth]{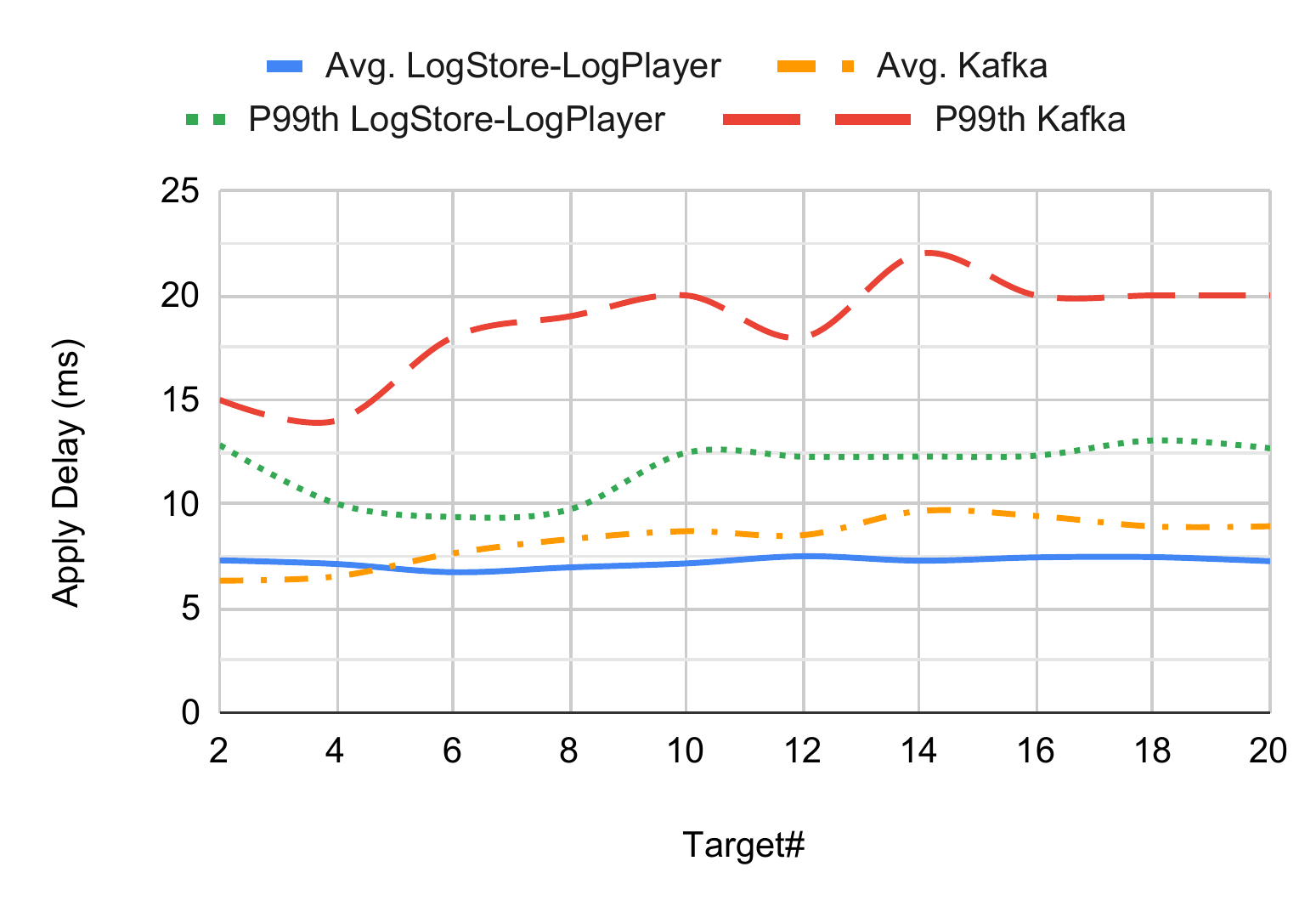}
}
  \subfigure[Apply Delay vs. Transaction Size \label{fig:applyDelay_size_kafka_nonT}]{\includegraphics[width=\diagramScale\columnwidth]{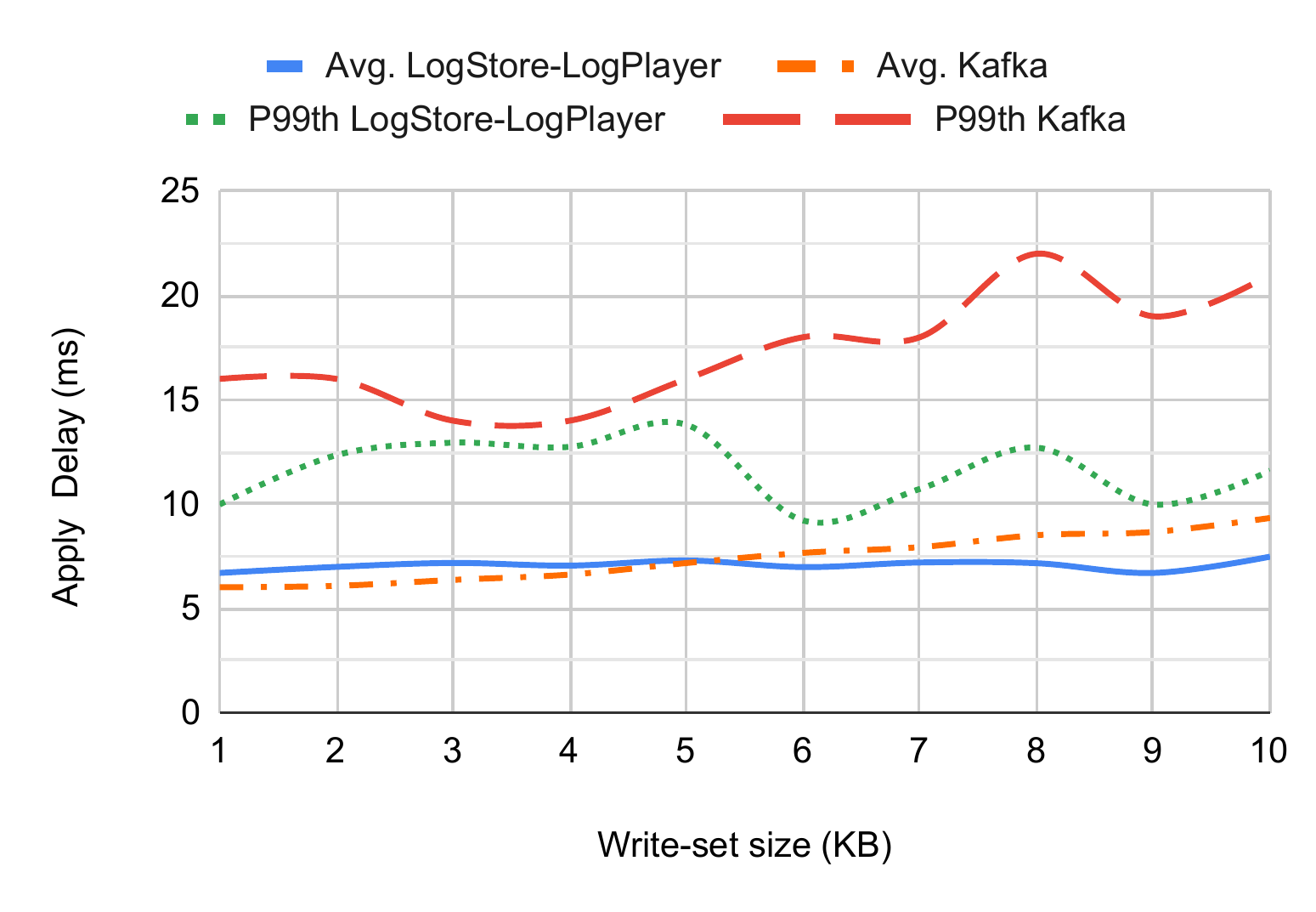}
}
\caption{Comparison with Kafka without exactly-once guarantee for different transaction sizes and number of targets}
\label{fig:applyDelay_kafka_nonT}
\end{figure*}

 Figure \ref{fig:applyDelay_kafka_nonT} compares the average and 99th percentile of apply delay of Kafka with that of LogStore-\name for various numbers of targets and transactions sizes when Kafka transactions are disabled. We set the size of transactions to 10 KB when we change the number of targets (Figure~\ref{fig:applyDelay_target_kafka_nonT}), and set the number of targets to 20 when we change the transaction size (Figure~\ref{fig:applyDelay_size_kafka_nonT}). Note that when transactions are disabled, Kafka cannot guarantee exactly-once semantics, but its performance is expected to be higher, as transaction overheads such as running 2PC are eliminated. Figure~\ref{fig:applyDelay_kafka_nonT} shows our system generally works better than Kafka even when Kafka does not satisfy exactly-once semantics and our system does, especially for larger numbers of targets and larger transaction sizes. The advantage of our system is more clear regarding the tail latency. For instance, for 20 targets and transaction size 10 KB, regarding the 99th percentile apply latency, our system is 45\% faster than Kafka. 

\begin{figure*}
  \centering
  \subfigure[Apply Delay vs. Target\# \label{fig:applyDelay_target_kafka}]{\includegraphics[width=\diagramScale\columnwidth]{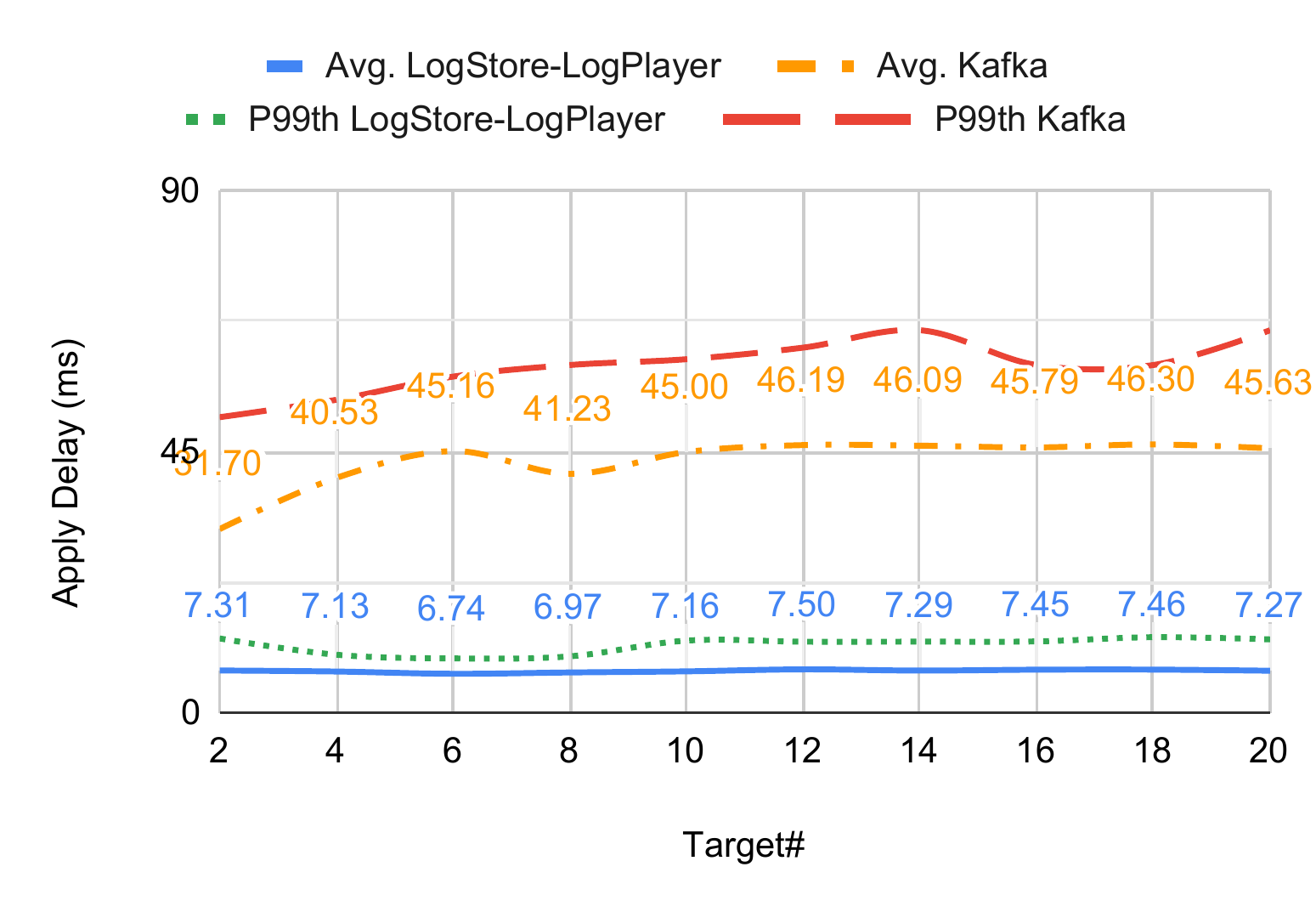}
}
  \subfigure[Apply Delay vs. Transaction Size \label{fig:applyDelay_size_kafka}]{\includegraphics[width=\diagramScale\columnwidth]{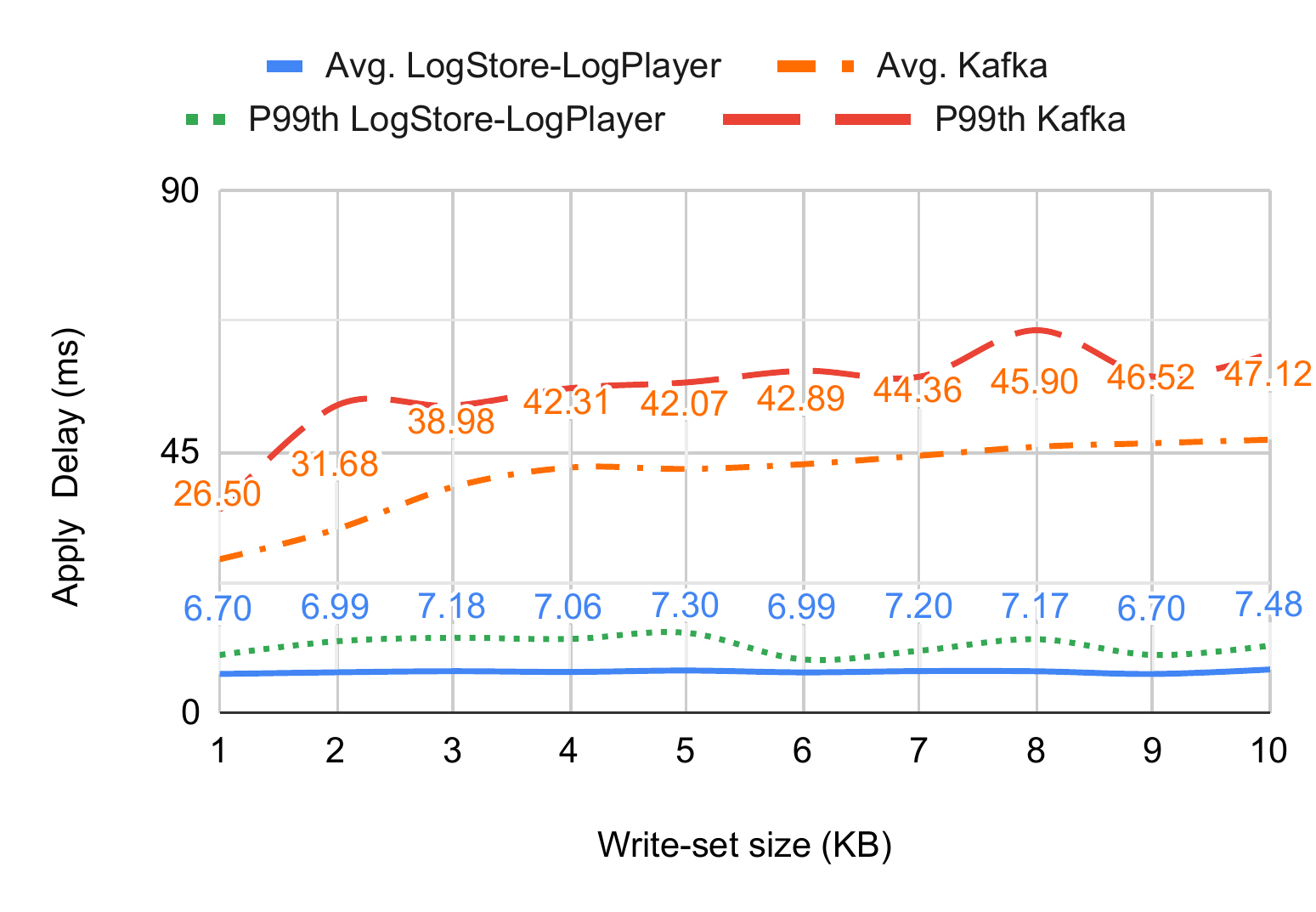}
}
\caption{Comparison with Kafka with exactly-once guarantee for different transaction sizes and number of targets}
\label{fig:applyDelay_kafa}
\end{figure*}

Now, we compare our system with Kafka when Kafka is configured for exactly-once semantics by enabling the transactions. As it is shown in Figure \ref{fig:applyDelay_kafa}, enabling Kafka transactions increases the apply delay significantly. This increase is resulted by the overhead of the 2PC protocol. 
In Figures \ref{fig:applyDelay_target_kafka}, and \ref{fig:applyDelay_size_kafka}, we presented the average apply delay values for Kafka and our system. As the number of targets increases, the apply delay of Kafka  increases almost linearly up to 10 targets. After 10 targets, the apply delay remains constant. That is due to the transaction size which is 10. Specifically, 10 keys can be hosted on at most 10 targets. Thus, adding more targets while keeping transactions size at 10 keys, does not significantly affect Kafka apply delay that is dominated by the 2PC overhead. For 10 targets, the average apply delay of Kafka is 45~ms, while the average delay of our system is only 7.16~ms which means, on average, our system is more than 4 times faster. The 99th percentile apply delay of Kafka is 61 ms, while it is only 12.47 ms for our system which means, regarding the tail latency, our system is more than 6 times faster. 

\section{Conclusion}
\label{sec:con}

In this paper, we presented the design of \name and our experience in using gRPC asynchronous streaming for delivering transactional mutations written to a log to backend storage shards. We model checked the correctness of \name using TLA+. Specifically, we proved \name is masking fault-tolerant for satisfying in-order exactly-once delivery of the log entries to the storage shards. TLA+ helped us find several bugs in our initial design that are fixed in the algorithms provided in this paper. We explained how \name can be configured with LogStore for a distributed database architecture. Our experimental results with the C++ implementation of the \name shows our design with gRPC asynchronous streaming provides efficient delivery of transactions to the storage shards. In all of our experiments, the median and average \name delay remained less than 1 millisecond. We showed that our system is significantly more efficient than existing streaming platforms such as Apache Kafka for designing a WAL system with the exactly-once guarantee. 

\balance
\bibliographystyle{ACM-Reference-Format}
\bibliography{bib}


\begin{thebibliography}{18}


\ifx \showCODEN    \undefined \def \showCODEN     #1{\unskip}     \fi
\ifx \showDOI      \undefined \def \showDOI       #1{#1}\fi
\ifx \showISBNx    \undefined \def \showISBNx     #1{\unskip}     \fi
\ifx \showISBNxiii \undefined \def \showISBNxiii  #1{\unskip}     \fi
\ifx \showISSN     \undefined \def \showISSN      #1{\unskip}     \fi
\ifx \showLCCN     \undefined \def \showLCCN      #1{\unskip}     \fi
\ifx \shownote     \undefined \def \shownote      #1{#1}          \fi
\ifx \showarticletitle \undefined \def \showarticletitle #1{#1}   \fi
\ifx \showURL      \undefined \def \showURL       {\relax}        \fi
\providecommand\bibfield[2]{#2}
\providecommand\bibinfo[2]{#2}
\providecommand\natexlab[1]{#1}
\providecommand\showeprint[2][]{arXiv:#2}

\bibitem[\protect\citeauthoryear{Ahn, Kang, Ren, Zhang, and Ben-Romdhane}{Ahn
  et~al\mbox{.}}{2019}]%
        {logstore}
\bibfield{author}{\bibinfo{person}{Jung-Sang Ahn}, \bibinfo{person}{Woon-Hak
  Kang}, \bibinfo{person}{Kun Ren}, \bibinfo{person}{Guogen Zhang}, {and}
  \bibinfo{person}{Sami Ben-Romdhane}.} \bibinfo{year}{2019}\natexlab{}.
\newblock \showarticletitle{Designing an efficient replicated log store with
  consensus protocol}. In \bibinfo{booktitle}{\emph{11th $\{$USENIX$\}$
  Workshop on Hot Topics in Cloud Computing (HotCloud 19)}}.
\newblock


\bibitem[\protect\citeauthoryear{Alpern and Schneider}{Alpern and
  Schneider}{1985}]%
        {liveness}
\bibfield{author}{\bibinfo{person}{B. Alpern} {and} \bibinfo{person}{F.~B.
  Schneider}.} \bibinfo{year}{1985}\natexlab{}.
\newblock \showarticletitle{Defining liveness}.
\newblock \bibinfo{journal}{\emph{Inform. Process. Lett.}}
  \bibinfo{volume}{21} (\bibinfo{year}{1985}), \bibinfo{pages}{181--185}.
\newblock


\bibitem[\protect\citeauthoryear{{Apache Flink}}{{Apache Flink}}{2019}]%
        {flink}
\bibfield{author}{\bibinfo{person}{{Apache Flink}}.}
  \bibinfo{year}{2019}\natexlab{}.
\newblock \bibinfo{title}{Apache Flink Homepage}.
\newblock \bibinfo{howpublished}{\url{https://flink.apache.org/}}.
\newblock


\bibitem[\protect\citeauthoryear{{Apache Kafka}}{{Apache Kafka}}{2019}]%
        {kafka}
\bibfield{author}{\bibinfo{person}{{Apache Kafka}}.}
  \bibinfo{year}{2019}\natexlab{}.
\newblock \bibinfo{title}{Apache Kafka Homepage}.
\newblock \bibinfo{howpublished}{\url{https://kafka.apache.org/}}.
\newblock


\bibitem[\protect\citeauthoryear{Arora and Gouda}{Arora and Gouda}{1993}]%
        {masking}
\bibfield{author}{\bibinfo{person}{A. Arora} {and} \bibinfo{person}{M.~G.
  Gouda}.} \bibinfo{year}{1993}\natexlab{}.
\newblock \showarticletitle{Closure and Convergence: {A} Foundation of
  Fault-Tolerant Computing}.
\newblock \bibinfo{journal}{\emph{{IEEE} Trans. Software Eng.}}
  \bibinfo{volume}{19}, \bibinfo{number}{11} (\bibinfo{year}{1993}),
  \bibinfo{pages}{1015--1027}.
\newblock


\bibitem[\protect\citeauthoryear{Chandy and Lamport}{Chandy and
  Lamport}{1985}]%
        {chandy}
\bibfield{author}{\bibinfo{person}{K~Mani Chandy} {and} \bibinfo{person}{Leslie
  Lamport}.} \bibinfo{year}{1985}\natexlab{}.
\newblock \showarticletitle{Distributed snapshots: Determining global states of
  distributed systems}.
\newblock \bibinfo{journal}{\emph{ACM Transactions on Computer Systems (TOCS)}}
  \bibinfo{volume}{3}, \bibinfo{number}{1} (\bibinfo{year}{1985}),
  \bibinfo{pages}{63--75}.
\newblock


\bibitem[\protect\citeauthoryear{{Confluent Tech Blog}}{{Confluent Tech
  Blog}}{2019}]%
        {kafkaBlog}
\bibfield{author}{\bibinfo{person}{{Confluent Tech Blog}}.}
  \bibinfo{year}{2019}\natexlab{}.
\newblock \bibinfo{title}{Transactions in Apache Kafka}.
\newblock
  \bibinfo{howpublished}{\url{https://www.confluent.io/blog/transactions-apache-kafka/}}.
\newblock


\bibitem[\protect\citeauthoryear{{Fauna}}{{Fauna}}{2019}]%
        {faunadb}
\bibfield{author}{\bibinfo{person}{{Fauna}}.} \bibinfo{year}{2019}\natexlab{}.
\newblock \bibinfo{title}{Fauna Homepage}.
\newblock \bibinfo{howpublished}{\url{https://fauna.com/}}.
\newblock


\bibitem[\protect\citeauthoryear{Gray and Reuter}{Gray and Reuter}{1992}]%
        {gray}
\bibfield{author}{\bibinfo{person}{Jim Gray} {and} \bibinfo{person}{Andreas
  Reuter}.} \bibinfo{year}{1992}\natexlab{}.
\newblock \bibinfo{booktitle}{\emph{Transaction processing: concepts and
  techniques}}.
\newblock \bibinfo{publisher}{Elsevier}.
\newblock


\bibitem[\protect\citeauthoryear{{gRPC}}{{gRPC}}{2019}]%
        {grpc}
\bibfield{author}{\bibinfo{person}{{gRPC}}.} \bibinfo{year}{2019}\natexlab{}.
\newblock \showarticletitle{gRPC Homepage}.
\newblock  (\bibinfo{year}{2019}).
\newblock
\newblock
\shownote{\url{https://grpc.io/}.}


\bibitem[\protect\citeauthoryear{{JanusGraph}}{{JanusGraph}}{2019}]%
        {janusgraph}
\bibfield{author}{\bibinfo{person}{{JanusGraph}}.}
  \bibinfo{year}{2019}\natexlab{}.
\newblock \bibinfo{title}{JanusGraph Homepage}.
\newblock \bibinfo{howpublished}{\url{https://janusgraph.org/}}.
\newblock


\bibitem[\protect\citeauthoryear{Kleppmann}{Kleppmann}{2017}]%
        {book}
\bibfield{author}{\bibinfo{person}{Martin Kleppmann}.}
  \bibinfo{year}{2017}\natexlab{}.
\newblock \bibinfo{booktitle}{\emph{Designing data-intensive applications: The
  big ideas behind reliable, scalable, and maintainable systems}}.
\newblock \bibinfo{publisher}{" O'Reilly Media, Inc."}.
\newblock


\bibitem[\protect\citeauthoryear{{Leslie Lamport}}{{Leslie Lamport}}{2019}]%
        {tla}
\bibfield{author}{\bibinfo{person}{{Leslie Lamport}}.}
  \bibinfo{year}{2019}\natexlab{}.
\newblock \bibinfo{title}{TLA+}.
\newblock
  \bibinfo{howpublished}{\url{https://lamport.azurewebsites.net/tla/tla.html}}.
\newblock


\bibitem[\protect\citeauthoryear{{NuRaft}}{{NuRaft}}{2019}]%
        {nuraft}
\bibfield{author}{\bibinfo{person}{{NuRaft}}.} \bibinfo{year}{2019}\natexlab{}.
\newblock \bibinfo{title}{NuRaft Github Repository}.
\newblock \bibinfo{howpublished}{\url{https://github.com/eBay/NuRaft}}.
\newblock


\bibitem[\protect\citeauthoryear{Ongaro and Ousterhout}{Ongaro and
  Ousterhout}{2014}]%
        {raft}
\bibfield{author}{\bibinfo{person}{Diego Ongaro} {and} \bibinfo{person}{John
  Ousterhout}.} \bibinfo{year}{2014}\natexlab{}.
\newblock \showarticletitle{In search of an understandable consensus
  algorithm}. In \bibinfo{booktitle}{\emph{2014 $\{$USENIX$\}$ Annual Technical
  Conference ($\{$USENIX$\}$$\{$ATC$\}$ 14)}}. \bibinfo{pages}{305--319}.
\newblock


\bibitem[\protect\citeauthoryear{Ren, Li, and Abadi}{Ren et~al\mbox{.}}{2019}]%
        {ren2019slog}
\bibfield{author}{\bibinfo{person}{Kun Ren}, \bibinfo{person}{Dennis Li}, {and}
  \bibinfo{person}{Daniel~J Abadi}.} \bibinfo{year}{2019}\natexlab{}.
\newblock \showarticletitle{SLOG: serializable, low-latency, geo-replicated
  transactions}.
\newblock \bibinfo{journal}{\emph{Proceedings of the VLDB Endowment}}
  \bibinfo{volume}{12}, \bibinfo{number}{11} (\bibinfo{year}{2019}),
  \bibinfo{pages}{1747--1761}.
\newblock


\bibitem[\protect\citeauthoryear{Thomson, Diamond, Weng, Ren, Shao, and
  Abadi}{Thomson et~al\mbox{.}}{2012}]%
        {thomson2012calvin}
\bibfield{author}{\bibinfo{person}{Alexander Thomson},
  \bibinfo{person}{Thaddeus Diamond}, \bibinfo{person}{Shu-Chun Weng},
  \bibinfo{person}{Kun Ren}, \bibinfo{person}{Philip Shao}, {and}
  \bibinfo{person}{Daniel~J Abadi}.} \bibinfo{year}{2012}\natexlab{}.
\newblock \showarticletitle{Calvin: fast distributed transactions for
  partitioned database systems}. In \bibinfo{booktitle}{\emph{Proceedings of
  the 2012 ACM SIGMOD International Conference on Management of Data}}. ACM,
  \bibinfo{pages}{1--12}.
\newblock


\bibitem[\protect\citeauthoryear{Zhang, Ren, Ahn, and Ben-Romdhane}{Zhang
  et~al\mbox{.}}{2019}]%
        {grit}
\bibfield{author}{\bibinfo{person}{Guogen Zhang}, \bibinfo{person}{Kun Ren},
  \bibinfo{person}{Jung-Sang Ahn}, {and} \bibinfo{person}{Sami Ben-Romdhane}.}
  \bibinfo{year}{2019}\natexlab{}.
\newblock \showarticletitle{{GRIT}: Consistent Distributed Transactions Across
  Polyglot Microservices with Multiple Databases}. In
  \bibinfo{booktitle}{\emph{2019 IEEE 35th International Conference on Data
  Engineering (ICDE)}}. IEEE, \bibinfo{pages}{2024--2027}.
\newblock


\end{thebibliography}

\end{document}